\documentclass[sn-mathphys-ay,referee,iicol]{sn-jnl}

\usepackage{graphicx}%
\usepackage{multirow}%
\usepackage{amsmath,amssymb,amsfonts}%
\usepackage{amsthm}%
\usepackage{mathrsfs}%
\usepackage[title]{appendix}%
\usepackage{xcolor}%
\usepackage{textcomp}%
\usepackage{manyfoot}%
\usepackage{booktabs}%
\usepackage{algorithm}%
\usepackage{algorithmicx}%
\usepackage{algpseudocode}%
\usepackage{listings}%
\usepackage{breqn}
\usepackage{cprotect}

\theoremstyle{thmstyleone}%
\theoremstyle{thmstyletwo}%

\theoremstyle{thmstylethree}%

\begin{document}

\title[SB-ETAS: using simulation based inference for scalable, likelihood-free inference for the ETAS model of earthquake occurrences]{SB-ETAS: using simulation based inference for scalable, likelihood-free inference for the ETAS model of earthquake occurrences }


\author*[1]{\fnm{Samuel} \sur{Stockman}}\email{sam.stockman@bristol.ac.uk}

\author[1]{\fnm{Daniel J.} \sur{Lawson}}\email{dan.lawson@bristol.ac.uk}

\author[2]{\fnm{Maximilian J.} \sur{Werner}}\email{max.werner@bristol.ac.uk}

\affil*[1]{\orgdiv{School of Mathematics}, \orgname{University of Bristol}, \orgaddress{\street{Woodland Rd}, \city{Bristol}, \postcode{BS8 1UG}, \country{United Kingdom}}}

\affil[2]{\orgdiv{School of Earth Sciences
}, \orgname{University of Bristol}, \orgaddress{\street{Wills Memorial Building}, \city{Bristol}, \postcode{BS8 1RL}, \country{United Kingdom}}}


\abstract{Performing Bayesian inference for the Epidemic-Type Aftershock Sequence (ETAS) model of earthquakes typically requires MCMC sampling using the likelihood function or estimating the latent branching structure. These tasks have computational complexity $O(n^2)$ with the number of earthquakes and therefore do not scale well with new enhanced catalogs, which can now contain an order of $10^6$ events. On the other hand, simulation from the ETAS model can be done more quickly $O(n \log n )$. We present SB-ETAS: simulation-based inference for the temporal ETAS model. This is an approximate Bayesian method which uses Sequential Neural Posterior Estimation (SNPE), a machine learning based algorithm for learning posterior distributions from simulations. SB-ETAS can successfully approximate ETAS posterior distributions on shorter catalogues where it is computationally feasible to compare with MCMC sampling. Compared with \texttt{inlabru}, another approximate bayesian method for the Hawkes process, SB-ETAS provides better coverage of the true posterior distribution. Furthermore, the scaling of SB-ETAS makes it feasible to fit to very large earthquake catalogs, such as one for Southern California dating back to 1981. SB-ETAS can find Bayesian estimates of ETAS parameters for this catalog in less than 10 hours on a standard laptop, which would have taken over 2 weeks using MCMC. Looking beyond the standard ETAS model, this simulation based framework would allow earthquake modellers to define and infer parameters for much more complex models that have intractable likelihood functions. The simulation-based framework readily extends to Hawkes-branching models in other applied domains beyond seismology. }

\maketitle
\section{Introduction}
The Epidemic Type Aftershock Sequence (ETAS) model \citep{ogata1988statistical} has been the most dominant way of modelling seismicity in both retrospective and fully prospective forecasting experiments \citep[e.g.][]{woessner2011retrospective, rhoades2018highlights,taroni2018prospective,cattania2018forecasting,mancini2019improving,mancini2020predictive, iturrieta2024evaluation} as well as in operational earthquake forecasting \citep{marzocchi2014establishment,rhoades2016retrospective,field2017synoptic,omi2019implementation}.
The model characterises the successive triggering of earthquakes, making it effective for forecasting aftershock sequences. The parameters of the model are also used by seismologists to characterise physical phenomena of different tectonic regions such as structural heterogeneity, stress, and temperature \citep[e.g.][]{utsu1995centenary} or relative plate velocity \citep[e.g.][]{ide2013proportionality}.

In recent years, an accelerated growth in the number of seismic sensors and machine learning algorithms for detecting the arrival times of earthquake phases \citep[e.g.][]{zhu2019phasenet}, has meant that earthquake catalogs have grown to a size that has overtaken what is computationally feasible to fit an ETAS model to. In California, a deployment of a dense network of seismic sensors over the last century combined with an active tectonic regime has resulted in a comprehensive dataset of earthquakes in the region \citep{hutton2010earthquake}. Furthermore, in more specific areas of California, through machine learning based seismic phase picking and template matching, enhanced earthquake catalogs have been created which contain many small previously undetected earthquakes \citep{white2019detailed, ross2019searching}. It is fair to assume that these datasets will only continue to grow in the future as past continuous data is reprocessed and future earthquakes are recorded. Determining whether increased data size results in improved earthquake forecasts is an important question for the seismological community. This work provides some methodology for answering that question by providing a simulation based estimation procedure for the ETAS model, with a scalability that make it feasible to infer ETAS parameters for such large datasets.


Most commonly, point estimates of ETAS parameters are found through maximum likelihood estimation (MLE). From these point estimates, forecasts can be issued by simulating multiple catalogs over the forecasting horizon. Forecast uncertainty is quantified by the distribution of simulations from MLE parameter values, however, this approach fails to quantify uncertainty contained in estimating the parameters themselves. Parameter uncertainty for MLE can be estimated using the Hessian of the likelihood \citep{ogata1978estimators,rathbun1996asymptotic,wang2010standard}, which requires a very large sample size to be effective, or through multiple runs of the MLE procedure with different initial conditions \citep{lombardi2015estimation}. 

Full characterisation of the parameter uncertainty is achieved with Bayesian inference, a procedure which returns the entire probability distribution over parameters conditioned on the observed data and updated from prior knowledge. This distribution over parameters, known as the posterior, does not have a closed form expression for the temporal and spatio-temporal ETAS model and so several approaches have used Markov Chain Monte Carlo (MCMC) to obtain samples from this distribution \citep{vargas2012bayesian,omi2015intermediate,ross2021bayesian,molkenthin2022gp}. These approaches evaluate the likelihood of the ETAS model during the procedure, which is a an operation with quadratic complexity $\mathcal{O}(n^2)$, and therefore are only suitable for catalogs up to 10,000 events. In fact, the GP-ETAS model by \cite{molkenthin2022gp} has cubic complexity $\mathcal{O}(n^3)$, since their spatially varying background rate uses a Gaussian-Process (GP) prior.

An alternative to MCMC is based on the Integrated Nested Laplace Approximation (INLA) method \citep{rue2017bayesian}, which approximates the marginal posterior distributions using latent Gaussian models. The implementation of INLA for the ETAS model, named \verb|inlabru|, by \cite{serafini2023approximation} demonstrated a factor 10 speed-up over the MCMC method for a catalog of 3,500 events but they do not provide results on larger catalogs. Section \ref{sec:BI} provides a more in depth overview of the MCMC and INLA methods.

Here we propose SB-ETAS: an alternative method for Bayesian inference for the ETAS model. SB-ETAS uses Simulation Based Inference (SBI), a family of approximate procedures which infer posterior distributions for parameters using simulations in place of the likelihood \citep{beaumont2002approximate,cranmer2020frontier}. Simulation from the ETAS model is $\mathcal{O}(n \log n)$, making it a procedure which scales better than the evaluation of the likelihood, $\mathcal{O}(n^2)$. Furthermore, specifying a model through simulation rather than the likelihood broadens the scope of available models to encompass greater complexity. From a suite of SBI methods developed recently using neural density estimators \citep{cranmer2020frontier}, we choose Sequential Neural Posterior Estimation (SNPE) since it does not involve any MCMC. SNPE trains a neural density estimator to approximate the posterior distribution from pairs of simulations and parameters. Section \ref{sec:SBI} provides an overview of SNPE and other methods of SBI.

SBI is typically used in the case where the likelihood function is intractable, for example when the likelihood involves integrating over many unobserved latent variables. However, in our use case the likelihood function is tractable, presenting an alternative potential for SBI for problems where the simulator scales better than the likelihood. Looking beyond the example presented in this study, the likelihood-free inference procedure that we present can be applied to other branching models for earthquakes, as well as to other Hawkes models outside seismology. Such models need only be defined as a forward simulator and wouldn't be constrained to have a likelihood function. For example, a large problem in earthquake modeling is the time varying detection of earthquakes, particularly following very large magnitude events  \citep[e.g.][]{hainzl2016rate,zhuang2017data}. These undetected events bias models such as ETAS that do not account for them \citep{seif2017estimating,stockman2023forecasting}. A branching model which deletes simulated events with a time varying detection rate does not result in a tractable likelihood but could still be fit to data using our procedure.

Other branching models for earthquakes do not have tractable likelihood functions. For example, the model from \cite{field2017spatiotemporal} integrates an ETAS model with a long-term geological fault recurrence rupture model, and lacks a well-defined likelihood function. This work provides a framework for estimating the parameters for such models that incorporate geological features, allowing for better calibration.

The remainder of this paper is structured as follows: In section \ref{sec:ETAS} we give an overview of the ETAS model along with existing procedures for performing Bayesian inference; section \ref{sec:SBI} gives an overview of SBI, following which we describe the details of SB-ETAS in section \ref{sec:SB-ETAS}. We present empirical results based on synthetic earthquake catalogs in section \ref{sec:E&R} and observational earthquake data from Southern California in section \ref{sec:SCEDC}, before finishing with a discussion in section \ref{sec:D&C}.
\section{The ETAS model}\label{sec:ETAS}
The temporal Epidemic Type Aftershock Sequence (ETAS) model \citep{ogata1988statistical} is a marked Hawkes process \citep{hawkes1971spectra} that describes the random times of earthquakes $t_i$ along with their magnitudes $m_i$ in the sequence $\textbf{x} = \{(t_1,m_1),(t_2,m_2),\dots,(t_n,m_n)  \} \in [0,T]^n \times \mathcal{M}^n \subset \mathbb{R}_+^n \times \mathcal{M}^n$. The natural ordering of events leads to the definition of a filtration $\mathcal{H}_t$ known as the history of the process, which contains information on all events up to time $t$. Marked Hawkes processes are usually specified by their conditional intensity function \citep{rasmussen2018lecture},
\begin{align}
    &\lambda(t,m|\mathcal{H}_t) =\\
    &\lim_{\Delta t, \Delta m \rightarrow 0} \frac{\mathbb{E}\left[N([t,t+\Delta t) \times B(m,\Delta m)|\mathcal{H}_t\right]}{\Delta t |B(m,\Delta m)|}.
\end{align}
where the counting measure $N(A)$ counts the events in the set $A \subset \mathbb{R}_+ \times \mathcal{M}$ and $|B(m,\Delta m)|$ is the Lebesgue measure of the ball $B(m,\Delta m)$ with radius $\Delta m$.\\
The ETAS model typically has the form,
\begin{equation}\label{eq:ETAS}
    \lambda(t,m|\mathcal{H}_t)  = \left(\mu + \sum_{i:t_i<t} g(t-t_i,m_i) \right)f_{GR}(m),
\end{equation}
where $\mu$ is a constant background rate of events, $g(t,m)$ is a non-negative excitation kernel which describes how past events contribute to the likelihood of future events and $f_{GR}(m)$ is the probability density of observing magnitude $m$. For the ETAS model this triggering kernel factorises the contribution from the magnitude and the time,
\begin{align}
    &g(t,m) =  k(m; K, \alpha) h(t;c,p)\\ 
    &k(m; K, \alpha)  = Ke^{\alpha(m-M_0)}\ :\ m \geq M_0\label{eq:utsu} \\ 
    &h(t;c,p) =  c^{p-1}(p-1)(t+c)^{-p}:\ t \geq 0 \label{eq:omori}
\end{align}
where the $k(m; K, \alpha)$ is known as the Utsu law of productivity \citep{utsu1970aftershocks} and $h(t;c,p)$ is a power law known as the Omori-Utsu decay \citep{utsu1995centenary}. The model comprises the five parameters ${\mu, K, \alpha, c, p}$. The magnitudes are said to be ``unpredictable" since they do not depend on previous events and are distributed according to the Gutenberg-Richter law for magnitudes \citep{gutenberg1936magnitude} with probability density $f_{GR}(m) = \beta e^{\beta(m-M_0)}$ on the support ${m : m \geq M_0}$.
\subsection{Branching Process Formulation}
An alternative way of formulating the ETAS model is as a Poisson cluster process \citep{rasmussen2013bayesian}. In this formulation a set of immigrants $I$ are realisations of a Poisson process with rate $\mu$. Each immigrant $t_i \in I$ has a magnitude $m_i$ with probability density $f_{GR}$ and generates offspring $S_i$ from an independent non-homogeneous Poisson process, with rate $g(t-t_i,m_i)$. Each offspring $t_j \in S_i$ also has magnitude $m_j$ with probability density $f_{GR}$ and generate offspring $S_j$ of their own. This process is repeated over generations until a generation with no offspring in time interval $[0,T]$ is produced. If the average number of offspring for a given event; $\frac{K\beta}{\beta -\alpha}$, is greater than one, the process is called super-critical and there is a non-zero probability that is continues infinitely. Although it is not observed in the data, this process is accompanied by latent branching variables $B = \{B_1, \dots, B_n \}$ which define the branching structure of the process,
\begin{align}
    &B_i = \begin{cases}
        0\text{ if } t_i \in I\ (\text{i.e. } i \text{ is a background event})\\
        j\text{ if } t_i \in S_j\  (\text{i.e. } i \text{ is an offspring of  } j)
    \end{cases}
\end{align}
Both the intensity function formulation as well as the branching formulation define different methods for simulating as well as inferring parameters for the Hawkes process. We now give a brief overview of some of these methods, and we direct the reader to \cite{reinhart2018review} for a more detailed review.
\subsection{Simulation}
A simulation algorithm based on the conditional intensity function was proposed by \cite{ogata1998space}. This algorithm requires generating events sequentially using a thinning procedure. Simulating forward from an event $t_i$, the time to the next event $\tau$ is proposed from a Poisson process with rate $\lambda(t_i|\mathcal{H}_{t_i})$. The proposed event $t_{i+1} = t_i + \tau $ is then rejected with probability $1- \frac{\lambda(t_i+\tau|\mathcal{H}_{t_i})}{\lambda(t_i|\mathcal{H}_{t_i})}$. This procedure is then repeated from the newest simulated event until a proposed event falls outside a predetermined time window $[0,T]$. \\
This procedure requires evaluating the intensity function $\lambda(t|\mathcal{H}(t))$ at least once for each one of $n$ events that are simulated. Evaluating the intensity function requires a summation over all events before time $t$, thus giving this simulation procedure time complexity $\mathcal{O}(n^2)$.\\~\\
Algorithm \ref{alg:hawkes_sim}, which instead simulates using the branching process formulation of the ETAS model was proposed by \cite{zhuang2004analyzing}. This algorithm simulates events over generations $G^{(i)}, i=0,\dots,$ until no more events fall within the interval $[0,T]$.
\begin{algorithm}
\caption{Hawkes Branching Process Simulation in the interval $[0,T]$}
\label{alg:hawkes_sim}
\begin{enumerate}
    \item Generate events $(t_i,m_i) \in G^{(0)}$ from a stationary Poisson process with intensity $\mu$ in $[0,T]$.
    \item $l=0$.
    \item For each $(t_i,m_i) \in G^{(l)}$ simulate its $N^{(i)}$ offspring, where $N^{(i)} \sim \text{Poisson}(k(m_i))$. Each offspring's time is generated from $h(t)$ and magnitude from $f_{GR}$ and are labelled $O_i^{(l)}$.
    \item $G^{(l+1)} = \bigcup_{i \in G^{(l)}} O_i^{(l)}$
    \item If $G^{(l)}$ is not empty, $l = l+1$ and return to step 3. 
    \item Sort and return $S = \bigcup_{j=0}^lG^{(j)}$ as the set of all simulated events. 
\end{enumerate}
\end{algorithm}\\
This procedure has time complexity $\mathcal{O}(n)$ for steps 1-5, since there is only a single pass over all events. An additional time constraint is added in step 6, where the whole set of events are sorted chronologically, which is at best $\mathcal{O}(n\log n)$.
\subsection{Bayesian Inference}\label{sec:BI}
Given we observe the sequence $\textbf{x}_{\text{obs}} = \{(t_1,m_1),(t_2,m_2),\dots,(t_n,m_n)  \}$ in the interval $[0,T]$, we are interested in the posterior probability $p(\theta|\textbf{x}_{\text{obs}})$ for the parameters $\theta = (\mu, K, \alpha,c,p)$ of the ETAS model defined in (\ref{eq:ETAS})-(\ref{eq:omori}), updated from some prior probability $p(\theta)$. The posterior distribution, expressed in Bayes' rule,
\begin{equation}
    p(\theta|\textbf{x}_{\text{obs}}) \propto p(\textbf{x}_{\text{obs}}|\theta)p(\theta),
\end{equation}
is known up to a constant of proportionality through the product of the prior $p(\theta)$ and the likelihood $p(\textbf{x}_{\text{obs}}|\theta)$, where,
\begin{align}
    \log p(\textbf{x}_{\text{obs}}|\theta) &= \sum_{i=1}^n \log\left[\mu + \sum_{j=1}^{i-1} h(t;c,p) k(m; K, \alpha) \right] \\
    &- \mu T - \sum_{i=1}^nk(m; K, \alpha)H(T-t_i; c, p).
\end{align}
Here, $H(t; c,p) = \int_0^t h(s;c,p)ds$, denotes the integral of the Omori decay kernel.\\
\cite{vargas2012bayesian} draw samples from the posterior $p(\theta|\textbf{x}_{\text{obs}})$ through independent random walk Markov Chain Monte Carlo (MCMC) with Metropolis-Hastings rejection of proposed samples. This approach, however, can suffer from slow convergence due to parameters of the ETAS model having high correlation \citep{ross2021bayesian}.\\
\cite{ross2021bayesian} developed an MCMC sampling scheme, \verb|bayesianETAS|, which conditions on the latent branching variables $B = \{B_1, \dots, B_n \}$. The scheme iteratively estimates the branching structure,
\begin{align}\label{eq:branching_matrix}
    &\mathbb{P}(B_i^{(k)} = j| \textbf{x}_{\text{obs}} , \theta^{(k-1)}) =\\ &\hspace{1cm} \begin{cases}
        \frac{\mu^{(k-1)}}{\mu^{(k-1)} + \sum_{j=1}^{i-1}k(m_j)h(t_i-t_j)} \ : j=0\\
        \frac{k(m_j)h(t_i-t_j)}{\mu^{(k-1)} + \sum_{j=1}^{i-1}k(m_j)h(t_i-t_j)}\ : j = 1,2,\dots,i-1
    \end{cases}
\end{align} 
and then samples parameters, $\theta^{(k)} = (\mu^{(k)},K^{(k)}, \alpha^{(k)},c^{(k)},p^{(k)})$, from the conditional likelihood,
\begin{align}
    &\log p(\textbf{x}_{\text{obs}}|\theta, B^{(k)}) = \\ &|S_0^{(k)}|\log \mu - \mu T
    + \sum_{j=1}^n \Biggl(-k(m_j;K,\alpha)H(T-t_j;c,p) \\ &+  |S_j^{(k)}| \log k(m_j;K,\alpha)  + \sum_{t_i \in S_j^{(k)}} \log h(t_i-t_j; c,p) \Biggr),
\end{align}
where $|S_j|$ denotes the number of events that were triggered by the event at $t_j$.\\
By conditioning on the branching structure, the dependence between parameters $(K,\alpha)$ and $(c,p)$ is reduced, decreasing the time it takes for the sampling scheme to converge. We can see from equation (\ref{eq:branching_matrix}) that estimating the branching structure from the data is a procedure that is $\mathcal{O}(n^2)$. Since for every event $i = 1,\dots,n$, to estimate its parent we must sum over $j = 1,\dots,i-1$. For truncated version of the time kernel $h(t)$, this operation can be streamlined to $\mathcal{O}(n)$. However, due to the heavy-tailed power-law kernel typically used, the complexity scaling remains high as significant truncation of the kernel is unfeasible.\\~\\
More recently \cite{serafini2023approximation} have constructed an approximate method of Bayesian inference for the ETAS model based on an Integrated Nested Laplace Approximation (INLA) implemented in the R-package \verb|inlabru| as well as a linear approximation of the likelihood. This approach expresses the log-likelihood as 3 terms,
\begin{align}
    &\log p(\textbf{x}_o | \theta) =\\ &\hspace{0.3cm} -\Lambda_0(\textbf{x}_{\text{obs}},\theta) - \sum_{i=1}^n \Lambda_i(\textbf{x}_{\text{obs}},\theta) + \sum_{i=1}^n \log \lambda(t_i|\mathcal{H}_{t_i}),
\end{align}
where, 
\begin{align*}
    \Lambda_0(\textbf{x}_{\text{obs}},\theta) &= \int_0^T\mu dt, \\ \Lambda_i(\textbf{x}_{\text{obs}},\theta) &= \sum_{h=1}^{C_i}\int_{b_{h,i}} g(t-t_i,m_i)dt\\ &= \sum_{h=1}^{C_i}\Lambda_i(\textbf{x}_{\text{obs}},\theta,b_{h,i}).
\end{align*}
where $b_{1,i},\dots,b_{C_i,i}$ are chosen to partition the interval $[t_{i-1},t_i]$.
The log-likelihood is then linearly approximated with a first order Taylor expansion with respect to the posterior mode $\theta^*$,
\begin{align*}
    &\widehat{\log p}(\textbf{x}_{\text{obs}}|\theta ; \theta^*) = -\widehat{\Lambda}_0(\textbf{x}_{\text{obs}},\theta,\theta^*) \\ &\hspace{0.1cm}- \sum_{i=1}^n\sum_{h=1}^{C_i}\widehat{\Lambda}_i(\textbf{x}_{\text{obs}},\theta,b_{h,i}; \theta^*) + \sum_{i=1}^n \widehat{\log \lambda}(\textbf{x}_{\text{obs}},\theta;\theta^*) \\
    & = -\exp\{\overline{\log{\Lambda}_0}(\textbf{x}_{\text{obs}},\theta,\theta^*)\} \\ &- \sum_{i=1}^n\sum_{h=1}^{C_i}\exp\{\overline{\log\Lambda}_i(\textbf{x}_{\text{obs}},\theta,b_{h,i}; \theta^*)\}\\ &+ \sum_{i=1}^n \overline{\log \lambda}(\textbf{x}_{\text{obs}},\theta;\theta^*),
\end{align*}
where the notation, $\widehat{\Lambda}$, denotes the approximation of $\Lambda$ and $\overline{\log \Lambda(  \ ;\theta^*)}$ denotes the first order Taylor expansion of $\log \Lambda$ about the point $\theta^*$.\\
The posterior mode $\theta^*$ is found through a Quasi-Newton optimisation method and the final posterior densities are found using INLA, which approximates the marginal posteriors $p(\theta_i|\textbf{x}_{\text{obs}})$ using a latent Gaussian model.\\
This approach speeds up computation of the posterior densities, since it only requires evaluation of the likelihood function during the search for the posterior mode. However, the approximation of the likelihood requires partitioning the space into a number of bins, which the authors recommend choosing as greater than 3 per observation. This results in the approximate likelihood having complexity $\mathcal{O}(n^2).$
\section{Simulation Based Inference}\label{sec:SBI}
A family of Bayesian inference methods have evolved from application settings in science, economics or engineering where stochastic models are used to describe complex phenomena. In this setting, the model may simulate data from a given set of input parameters, however, the likelihood of observing data given parameters is intractable. The task in this setting is to approximate the posterior $p(\theta|\textbf{x}_{\text{obs}}) \propto p(\textbf{x}_{\text{obs}}|\theta)p(\theta)$, with the restriction that we cannot evaluate $p(\textbf{x}|\theta)$ but we have access to the likelihood implicitly through samples $\textbf{x}_r \sim p(\textbf{x}|\theta_r)
$ from a simulator, for $r = 1,\dots,R$ and where $\theta_r \sim p(\theta)$. This approach is commonly referred to as Simulation Based Inference (SBI) or likelihood-free inference.\\
Until recently, the predominant approach for SBI was Approximate Bayesian Computation \citep{beaumont2002approximate}. In its simplest form, parameters are chosen from the prior $\theta_r \sim p(\theta),\ r=1,\dots,R$, the simulator then generates samples $\textbf{x}_r \sim p(\textbf{x}|\theta_r),\ r=1,\dots,R$, and each sample is kept if it is within some tolerance $\epsilon$ of the observed data, i.e. $d(\textbf{x}_r,\textbf{x}_{\text{obs}}) < \epsilon$ for a given distance function $d(\cdot,\cdot)$. \\
This approach, although exact when $\epsilon \rightarrow 0$, is inefficient with the use of simulations. Sufficiently small $\epsilon$ requires simulating an impractical number of times, and this issue scales poorly with the dimension of $\textbf{x}$. In light of this an MCMC approach to ABC makes proposals for new simulator parameters $\theta_r \sim q(\cdot |\theta_{r-1})$ using a Metropolis-Hastings kernel \citep{beaumont2002approximate,marjoram2003markov}. This leads to a far higher acceptance of proposed simulator parameters but still scales poorly with the dimension of $\textbf{x}$.\\
In order to cope with high dimensional simulator outputs $\textbf{x} \in \mathbb{R}^n$, summary statistics $S(\textbf{x})\in \mathbb{R}^d$ are chosen to reduce the dimension of the sample whilst still retaining as much information as possible. These are often chosen from domain knowledge or can be learnt as part of the inference procedure \citep{prangle2014semi}. Summary statistics $S(\textbf{x})$ are then used in place of $\textbf{x}$ in any of the described methods for SBI.
\subsection{Neural Density Estimation}
\begin{figure}
    \centering
    \includegraphics[width= 0.5\textwidth]{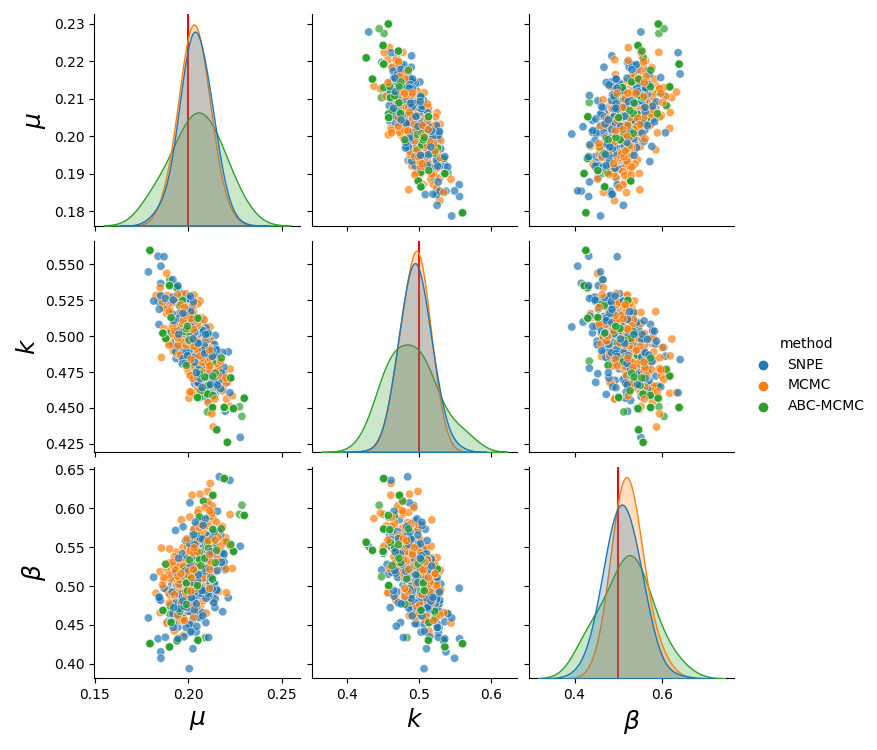}
    \caption{Posterior densities for a univariate Hawkes process with exponential kernel. The `observed' data contains 4806 events and was simulated from parameters indicated in red on the diagonal plots. In green are posterior samples found using the ABC-MCMC method for SBI using 300,000 simulations. In blue are posterior samples from SNPE using the same summary statistics as ABC-MCMC but only 10,000 simulations. In orange are posterior samples found using MCMC sampling with likelihood function. A $\text{Uniform}([0.05, 0,0],[0.85, 0.9,3])$ prior was used for all three methods.}
    \label{fig:ABC-HAWKES}
\end{figure}

Recently, SBI has seen more rapid development as a result of neural network based density estimators \citep{papamakarios2016fast,papamakarios2017masked,lueckmann2017flexible}, which seek to approximate the density $p(x)$ given samples of points $x \sim p(x)$. A popular method for neural density estimation is normalising flows \citep{rezende2015variational}, in which a neural network parameterizes an invertible transformation $x = g_\phi(u)$, of a variable $u$ from a simple base distribution $p(u)$ into the target distribution of interest. In practice, the transformation is typically composed of a stack of invertible transformations, which allows it to learn the complex target density. The parameters of the transformation are trained through maximising the likelihood of observing $p_g(x)$, which is given by the change of variables formula. Since $x$ is expressed as a transformation of a simple distribution $u \sim p(u)$, samples from the learnt distribution $p_g(x)$ can be generated by sampling from $p(u)$ and passing the samples through the transformation. Neural density estimators may also be generalised to learn conditional densities $p(\textbf{x}|\textbf{y})$ by conditioning the transformation $g_\phi$ on the variable $y$ \citep{papamakarios2017masked}.\\
In the task of SBI, a neural density estimator can be trained on pairs of samples $\theta_r \sim p(\theta), \textbf{x}_r \sim p(\textbf{x}|\theta_r)$ to approximate either the likelihood $p(\textbf{x}_{\text{obs}}|\theta)$ or the posterior density $p(\theta|\textbf{x}_{\text{obs}})$, from which posterior samples can be obtained. If the posterior density is estimated, in a procedure known as Neural Posterior Estimation (NPE) \citep{lueckmann2017flexible}, then samples can be drawn from the normalising flow. If the likelihood is estimated, known as Neural Likelihood Estimation (NLE) \citep{papamakarios2019sequential}, then the approximate likelihood can be used in place of the true likelihood in a MCMC sampling algorithm to obtain posterior samples. These density estimation techniques generally outperform ABC-based methods since they are able to efficiently interpolate between different simulations \citep[Figure \ref{fig:ABC-HAWKES},][]{lueckmann2021benchmarking}. Other neural network methods exist for SBI such as ratio estimation \citep{izbicki2014high} or score matching \citep{geffner2022score, sharrock2022sequential}, however, we direct the reader to \citep{cranmer2020frontier} for a more comprehensive review of modern SBI.

\section{SB-ETAS}\label{sec:SB-ETAS}
\begin{figure*}
    \centering
    \includegraphics[width =\textwidth]{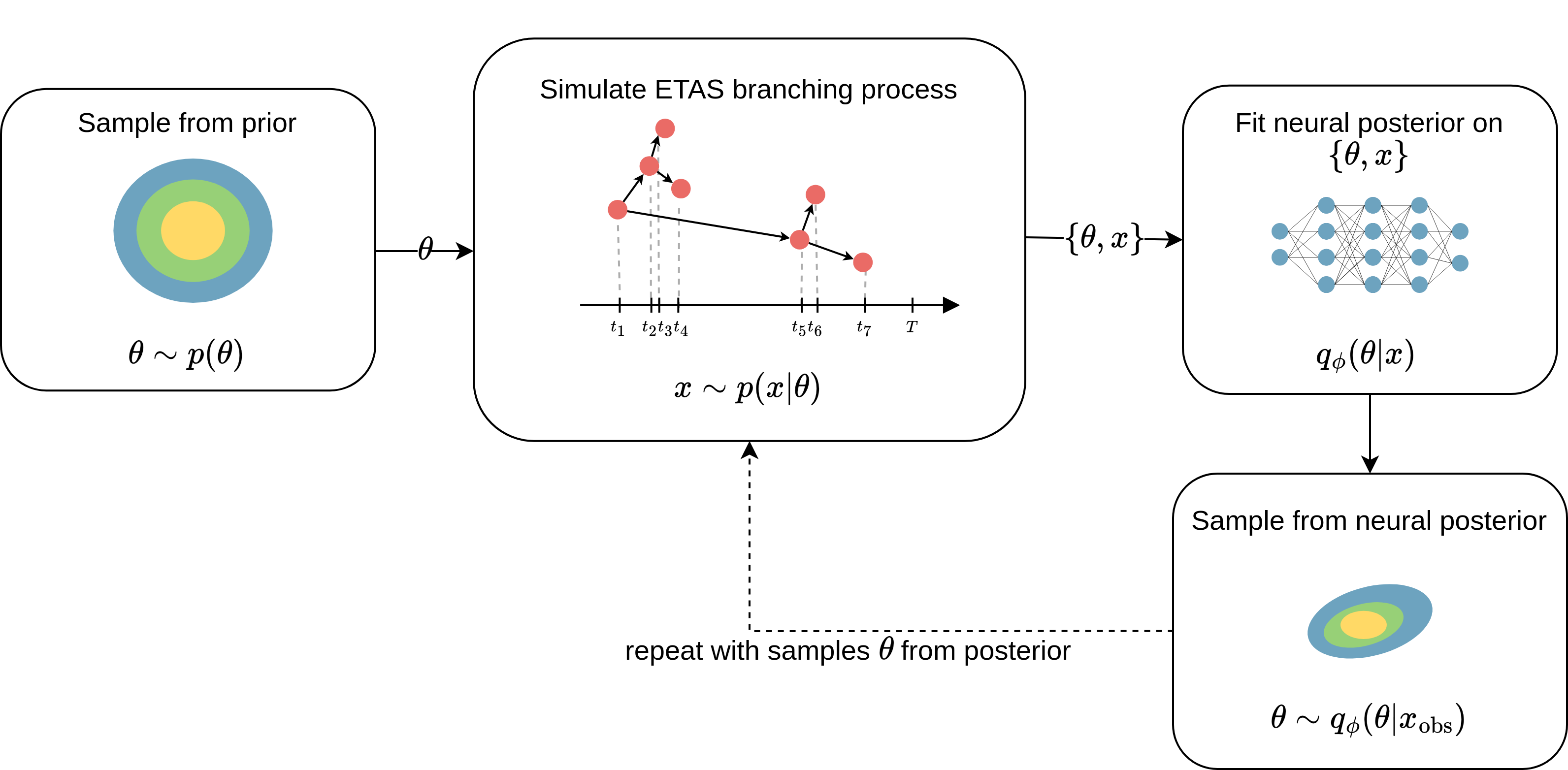}
    \caption{An outline of the SB-ETAS inference procedure. Samples from the prior distribution are used to simulate many ETAS sequences. A neural density estimator is then trained on the parameters and simulator outputs to approximate the posterior distribution. Samples from the posterior given the observed earthquake sequence can then be used to improve the estimate over rounds or are returned as the final posterior samples.}
    \label{fig:schematic}
\end{figure*}
We now present SB-ETAS, our scalable method for Bayesian inference for the ETAS model. The method avoids having to evaluate the computationally expensive likelihood function and instead leverages fast simulation from the ETAS branching process. The inference method uses Sequential Neural Posterior Estimation (SNPE) \citep{papamakarios2016fast,lueckmann2017flexible}, a modified version of NPE which performs inference over rounds. Each round, an estimate of the posterior proposes new samples for the simulator, a neural density estimator is trained on those samples and the estimated posterior is updated (Figure \ref{fig:schematic}, Algorithm \ref{alg:SB-ETAS}). SNPE was chosen over other methods of Neural-SBI, as it avoids the need to perform MCMC sampling, a slow procedure. Instead, sampling from the posterior is fast since the approximate posterior is a normalising flow.

\begin{algorithm}[t]
 \caption{SB-ETAS}\label{alg:SB-ETAS}

  \textbf{Input:} observed data $\textbf{x}_{\text{obs}}$, summary statistic $S(\textbf{x})$, estimator $q_\phi(\theta|\textbf{x})$, prior $p(\theta)$, number of rounds $K$, simulations per round $L$.\\

 \vspace{-1em}
\begin{algorithmic}
   \State $q^{0}_\phi(\theta|\textbf{x}_{\text{obs}}) = p(\theta)$ and $\mathcal{D} =\{\}$
  \For{$k = 1:K$}
  \For{$l=1:L$}
  \State sample $\theta_l \sim q^{k-1}_\phi(\theta|\textbf{x}_{\text{obs}})$
  \State simulate $\textbf{x}_l \sim p(\textbf{x}|\theta_l)$ using Algorithm \ref{alg:hawkes_sim}
  \State add $(\theta_l,S(\textbf{x}_l))$ to $\mathcal{D}$
  \EndFor
  \State train $q_\phi^{k}(\theta|\textbf{x}_{\text{obs}})$ on $\mathcal{D}$
  \EndFor
\end{algorithmic}
 \begin{flushleft}
  \textbf{Output:} approximate posterior $\hat{p}(\theta|\textbf{x}_{\text{obs}}) = \hat{q}_\phi^{K}(\theta|\textbf{x}_{\text{obs}})$
 \end{flushleft}
\end{algorithm}
\subsection{Summary Statistics}
What differentiates Hawkes process models from other simulator models used in Neural-SBI is that the output of the simulator $\textbf{x} = (t_1,m_1),\dots,(t_n,m_n)$ itself has random dimension. For a specified time interval over which to simulate earthquakes $[0,T]$, one particular parameter $\theta_1$ will generate different numbers of events if simulated repeatedly. This is problematic for neural density estimators since even though they are successful over high dimensional data, they require a fixed dimensional input. For this reason, we fix the dimension of simulator output through calculating summary statistics of the data $S(\textbf{x})$. \\
Works to perform ABC on the univariate Hawkes process with exponential decay kernel have found summary statistics that perform well in that setting. \cite{ertekin2015reactive} use the histogram of inter-event times as summary statistics as well as the number of events. \cite{deutsch2021abc} extend these summary statistics by adding Ripley's K statistic \citep{ripley1977modelling}, which is a popular choice of summary statistic in spatial point processes \citep{moller2003introduction}. Figure \ref{fig:ABC-HAWKES} shows the performance of the ABC-MCMC method developed by \cite{deutsch2021abc}, using the aforementioned summary statistics. Using SNPE on the same summary statistics yields a more confident estimation of the ``true'' posterior which is found through MCMC sampling using the likelihood and requires far fewer simulations (10,000 versus 300,000).\\
The ETAS model is more complex than a univariate Hawkes process since it is both marked (i.e. it contains earthquake magnitudes) and contains a power law decay kernel which decays much more slowly than exponential, making it harder to estimate \citep{bacry2016estimation}. For SB-ETAS we borrow similar summary statistics to \cite{ertekin2015reactive}, namely $S_1(\textbf{x}) = \log( \# \text{ events})$, $S_2,\dots,S_4(\textbf{x}) = $20th, 50th and 90th quantiles of the inter-event time histogram. Similar to \cite{deutsch2021abc}, we use another statistic $S_5(\textbf{x})$, which is the ratio of the mean and median of the inter-event time histogram.
\subsubsection{Ripley's K Statistic}
For the remaining summary statistics, we develop upon the introduction of Ripley's K statistic by \citep{deutsch2021abc}. For a univariate point process $\textbf{x} = (t_1,\dots,t_n)$, Ripley's K statistic is \citep{dixon2001ripley},
\begin{align}
    &K(\textbf{x},w) = \\ &\frac{1}{\lambda}\mathbb{E}(\# \text{ of events within }w\text{ of a random event} ).
\end{align}
Here, $\lambda$ is the unconditional rate of events in the time window $[0,T]$.  An estimator for the K-statistic is derived by \cite{diggle1985kernel},
\begin{equation}
    \hat{K}(\textbf{x},w) = \frac{T}{n^2}\sum_{i=1}^n \sum_{j\neq i}\mathbb{I}(0< t_j-t_i \leq w).
\end{equation}
Despite containing a double-sum, computation of this estimator has complexity $\mathcal{O}(n)$ since $\{t_i\}_{i=1}^n$ is an ordered sequence, i.e. $(t_3 - t_1 < w) \Rightarrow (t_3 - t_2 < w)$. Calculation of Ripley K-statistic therefore satisfies the complexity requirement of our procedure if the number of windows $w$ for which we evaluate $\hat{K}(\textbf{x},w)$ is less than $\log n$. In fact, our results suggest that less than 20 are required.\\
The use of Ripley's K-statistic for non-marked Hawkes data is motivated by \citep{bacry2014second}, who show that second-order properties fully characterise a Hawkes process and can be used to estimate a non-parametric triggering kernel. \cite{bacry2016estimation} go on to give a recommendation for a binning strategy to estimate slow decay kernels such as a power law, using a combination of linear and log-scaling. It therefore seems reasonable to define $S_6(\textbf{x})\dots S_{23}(\textbf{x})= \hat{K}(\textbf{x},w)$, where $w$ scales logarithmically between $[0,1]$ and linearly above $1$.\\
We modify Ripley's K-statistic to account for the particular interaction between marks and points in the ETAS model. Namely, the magnitude of an earthquake directly affects the clustering that occurs following it, expressed in the productivity relationship (\ref{eq:utsu}). In light of this, we define a magnitude thresholded Ripley K-statistic,
\begin{align}
    &K_T(\textbf{x},w,M_T) = \\ &\frac{1}{\lambda_{T}}\mathbb{E}(\# \text{ events within }w\text{ of an event }m_i\geq M_T),
\end{align}
where $\lambda_T$ is the unconditional rate of events above $M_T$. One can see that $K_T(\textbf{x},w,M_0) = K(\textbf{x},w) $. We estimate $K_T$ with
\begin{equation}
        \hat{K}_T(\textbf{x},w,M_T) = \frac{T}{\nu^2}\sum_{i:m_i\geq M_T} \sum_{j\neq i}\mathbb{I}(0< t_i-t_j \leq w),
\end{equation}
where $\nu$ is the number of events above magnitude threshold $M_T$. For general $M_T$, we lose the $\mathcal{O}(n)$ complexity that the previous statistic has, instead it is $\mathcal{O}(\nu n)$. However, if the threshold is chosen to be large enough, evaluation of this estimator is fast. In our experiments, $M_T$ is chosen to be $(4.5,5,5.5,6)$, with $w = (0.2,0.5,1,3)$. This defines the remaining statistics $S_{24}(\textbf{x}),\dots,S_{39}(\textbf{x})$.

\section{Experiments and Results}\label{sec:E&R}
To evaluate the performance of SB-ETAS, we conduct inference experiments on a series of synthetic ETAS catalogs. On each simulated catalog we seek to obtain 5000 samples from the posterior distribution of ETAS parameters. The latent variable MCMC inference procedure, \verb|bayesianETAS|, will be used as a reference model in our experiments since it uses the ETAS likelihood without making any approximations. We compare samples from this exact method with samples from approximate methods, \verb|inlabru| and SB-ETAS.

Multiple catalogs are simulated from a fixed set of ETAS parameters, $(\mu,k,\alpha,c,p) = (0.2,0.2,1.5,0.5,2)$ with magnitude of completeness $M_0 = 3$ and Gutenberg-Richter distribution parameter $\beta = 2.4$. Each new catalog is simulated in a time window $[0,T]$, where $T \in (10,20,30,40,50,60,70,80,90,100,250,500,1000) \times  10^3$. 

\begin{figure}
    \centering
    \includegraphics[width=0.5\textwidth]{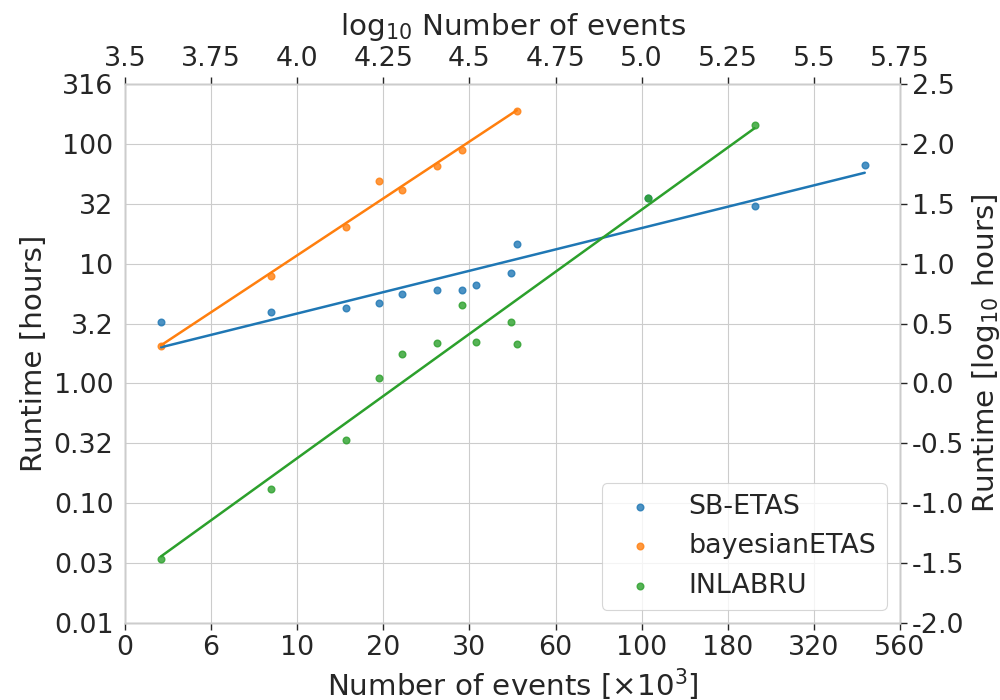}
    \caption{The runtime for parameter inference versus the catalog size for SB-ETAS, \texttt{inlabru} and \texttt{bayesianETAS}. Separate ETAS catalogs were generated with the same intensity function parameters but for varying size time-windows. The runtime in hours and the number of events are plotted in log-log space.}
    \label{fig:runtime}
\end{figure}
\begin{figure*}[h]
    \centering
    \includegraphics[width=\textwidth]{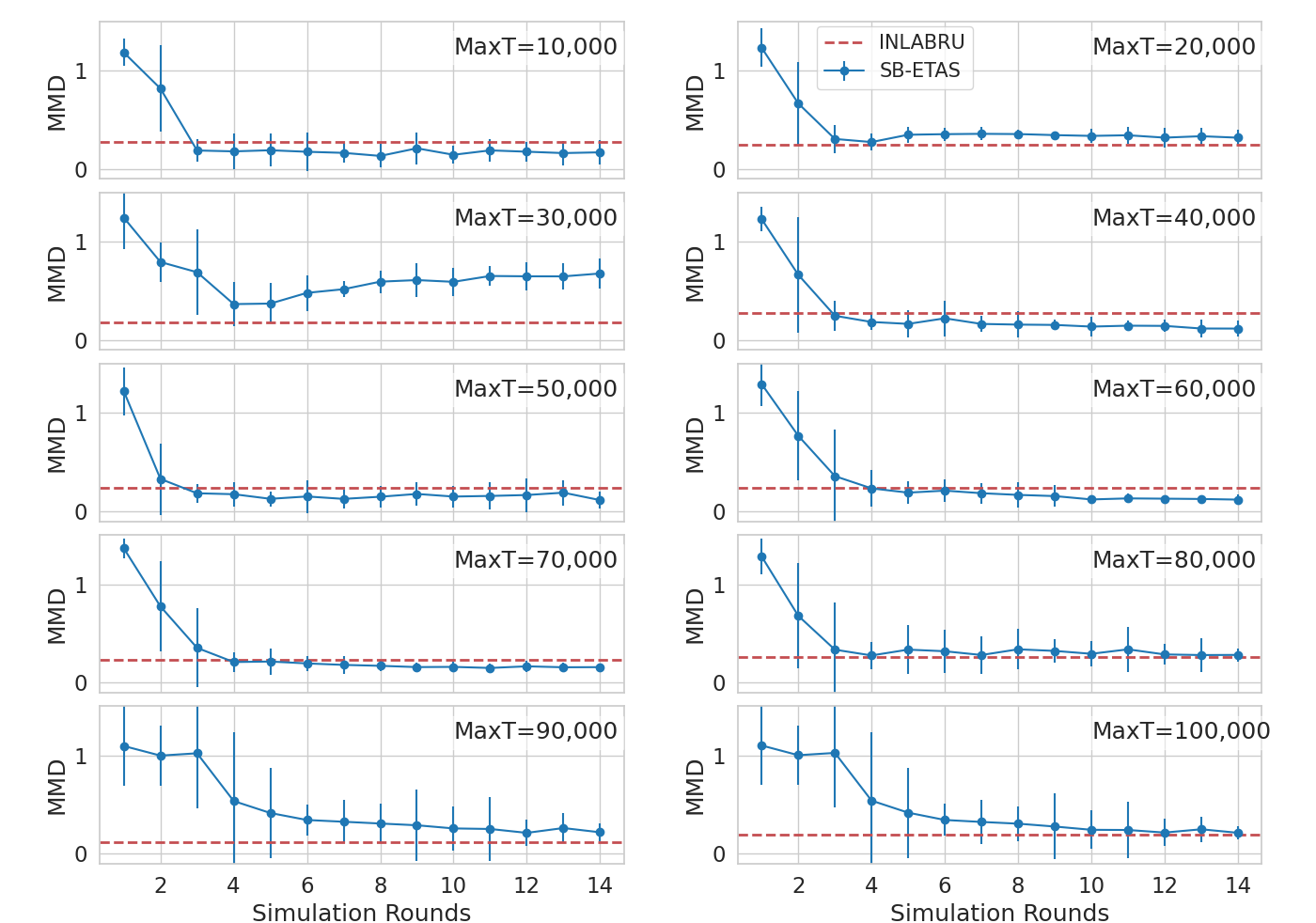}
    \caption{Maximum Mean Discrepancy for samples from each round of simulations in SB-ETAS. Each plot corresponds to a different simulated ETAS catalog simulated with identical model parameters but over a different length time-window (MaxT). In red is the performance metric evaluated for samples from \texttt{inlabru}. 95\% confidence intervals are plotted for SB-ETAS across 10 different initial seeds.}
    \label{fig:mmd}
\end{figure*}
\begin{figure*}[h]
    \centering
    \includegraphics[width=\textwidth]{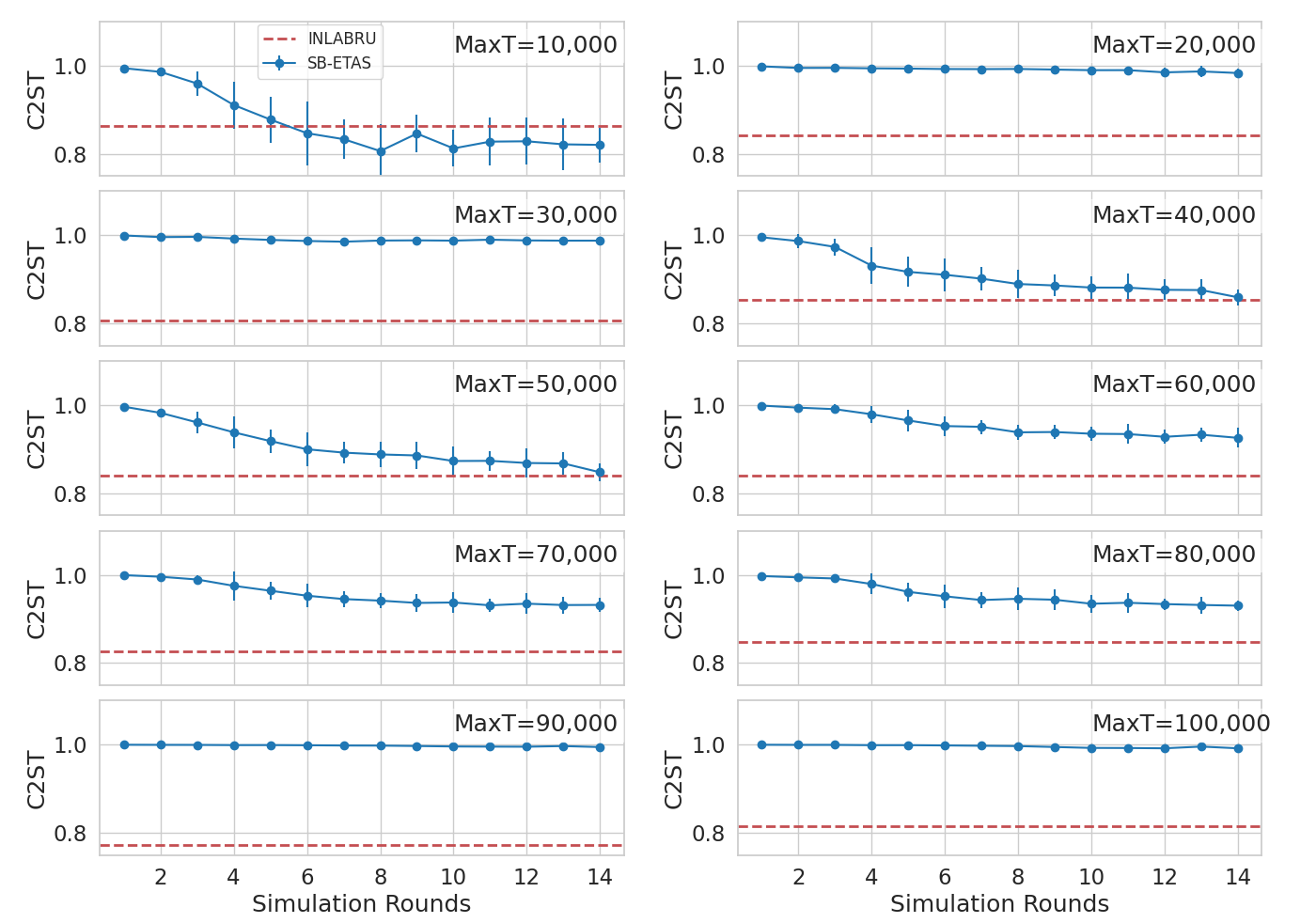}
    \caption{Classifier Two-Sample Test scores for samples from each round of simulations in SB-ETAS. Each plot corresponds to a different simulated ETAS catalog simulated with identical model parameters but over a different length time-window (MaxT). In red is the performance metric evaluated for samples from \texttt{inlabru}. 95\% confidence intervals are plotted for SB-ETAS across 10 different initial seeds.}
    \label{fig:c2st}
\end{figure*}
Figure \ref{fig:runtime} shows the runtime of each inference method as a function of the number of events in each catalog. Each method was run on a high-performance computing node with eight 2.4 GHz Intel E5-2680 v4 (Broadwell) CPUs, which is equivalent to what is commonly available on a standard laptop. On the catalogs with up to 100,000 events, \verb|inlabru| is the fastest inference method, around ten times quicker on a catalog of 20,000 events. However, the superior scaling of SB-ETAS allows it to be run on the catalog of $\sim 500,000$ events, which was unfeasible for \verb|inlabru| given the same computational resources i.e. it exceeded a two week time limit. The gradient of 2 for both \verb|bayesianETAS| and \verb|inlabru| in log-log space confirm the $\mathcal{O}(n^2)$ time complexity of both inference methods. SB-ETAS, on the other hand, has gradient $\frac{2}{3}$ which suggests that the theoretical $\mathcal{O}(n\log n)$ time complexity is a conservative upper-bound.

The prior distributions for each implementation are not identical since each has its own requirements. Priors are chosen to replicate the fixed implementation in the \verb|bayesianETAS| package,
\begin{align}
    \mu &\sim \text{Gamma}(0.1,0.1)\\
    K,\alpha,c &\sim \text{Unif}(0,10)\\
    p &\sim \text{Unif}(1,10).
\end{align}
The implementation of \verb|inlabru| uses a transformation $K_b = \frac{K(p-1)}{c}$, with prior $K_b \sim \text{Log-Normal}(-1,2.03)$ chosen by matching $1\%$ and $99\%$ quantiles with the \verb|bayesianETAS| prior for $K$. SB-ETAS uses a $\mu \sim \text{Unif}(0.05,0.3)$ prior in place of the gamma prior as well as enforcing a sub-critical parameter region $K\beta < \beta-\alpha$ \citep{zhuang2012basic}. Both the uniform prior and the restriction on $K$ and $\alpha$ stop unnecessarily long or infinite simulations.

\begin{figure}[h!]
    \hspace{-0.5cm}\includegraphics[width=0.5\textwidth]{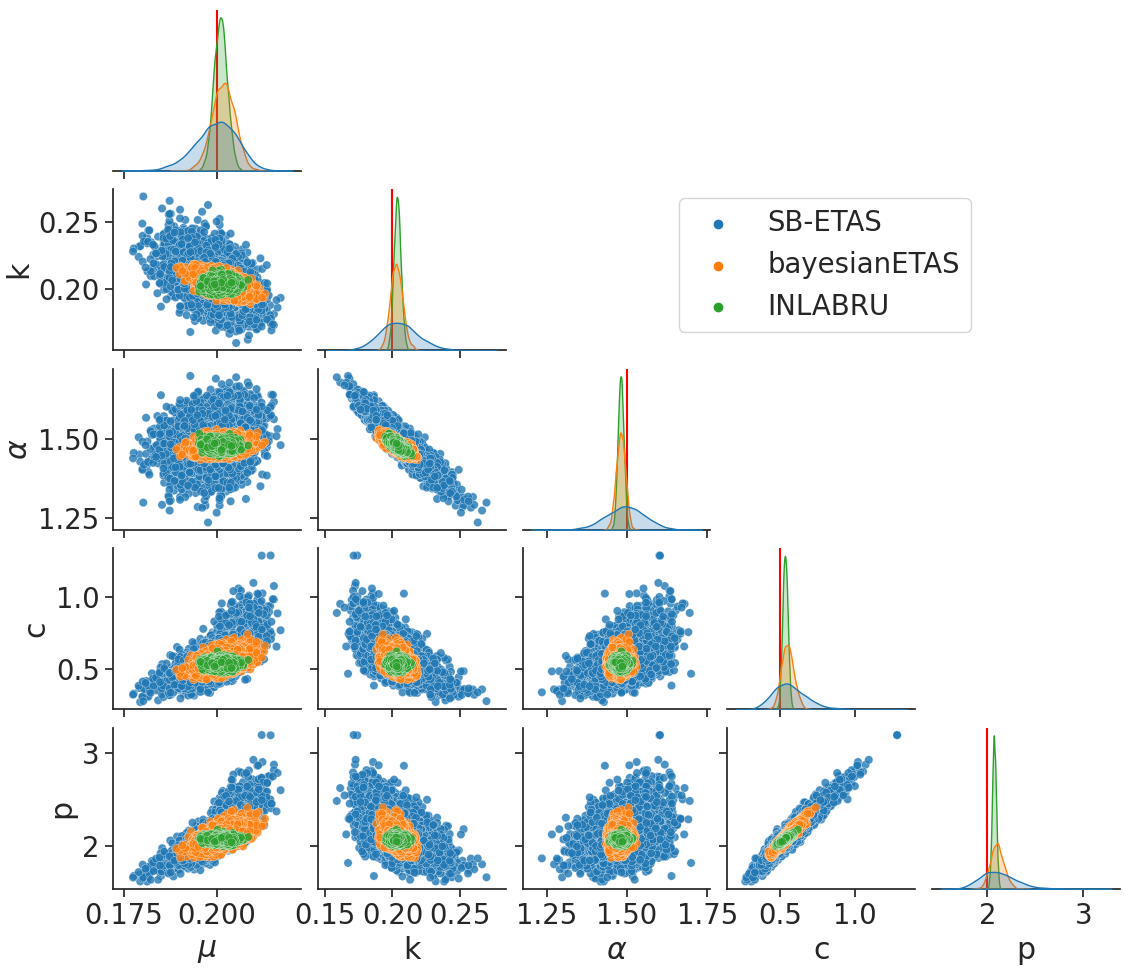}
    \caption{Samples from the posterior distribution of ETAS parameters for the simulated catalog with $T=60,000$, for \texttt{bayesianETAS}, \texttt{inlabru} and SB-ETAS. The data generating parameters are marked in red in the diagonal plots.}
    \label{fig:samples}
\end{figure}

Once samples are obtained from SB-ETAS and \verb|inlabru|, we measure their (dis)similarity with samples from the exact method \verb|bayesianETAS| using the Maximum Mean Discrepancy (MMD) \citep{gretton2012kernel} and the Classifier Two-Sample Test (C2ST) \citep{lehmann2005testing,lopez2016revisiting}. Figures \ref{fig:mmd} and \ref{fig:c2st} show the values of these performance metrics for samples from each of 15 rounds of simulations in SB-ETAS compared with the performance of \verb|inlabru|. Since SB-ETAS involves random sampling in the procedure, we repeat it across 10 different seeds and plot a 95\% confidence intervals. In general across the 10 synthetic catalogs, SB-ETAS and \verb|inlabru| are comparable in terms of MMD (Figure \ref{fig:mmd}) and \verb|inlabru| performs best in terms of C2ST (Figure \ref{fig:c2st}). Figure \ref{fig:samples} shows samples from the $T=60,000$. Samples from \verb|inlabru| are overconfident with respect to the \verb|bayesianETAS| samples, whereas SB-ETAS samples are more conservative. This phenomenon is shared across the samples from all the simulated catalogs and we speculate that it accounts for the difference between the two metrics.

\begin{figure*}[ht!]
    \centering
    \includegraphics[width=\textwidth]{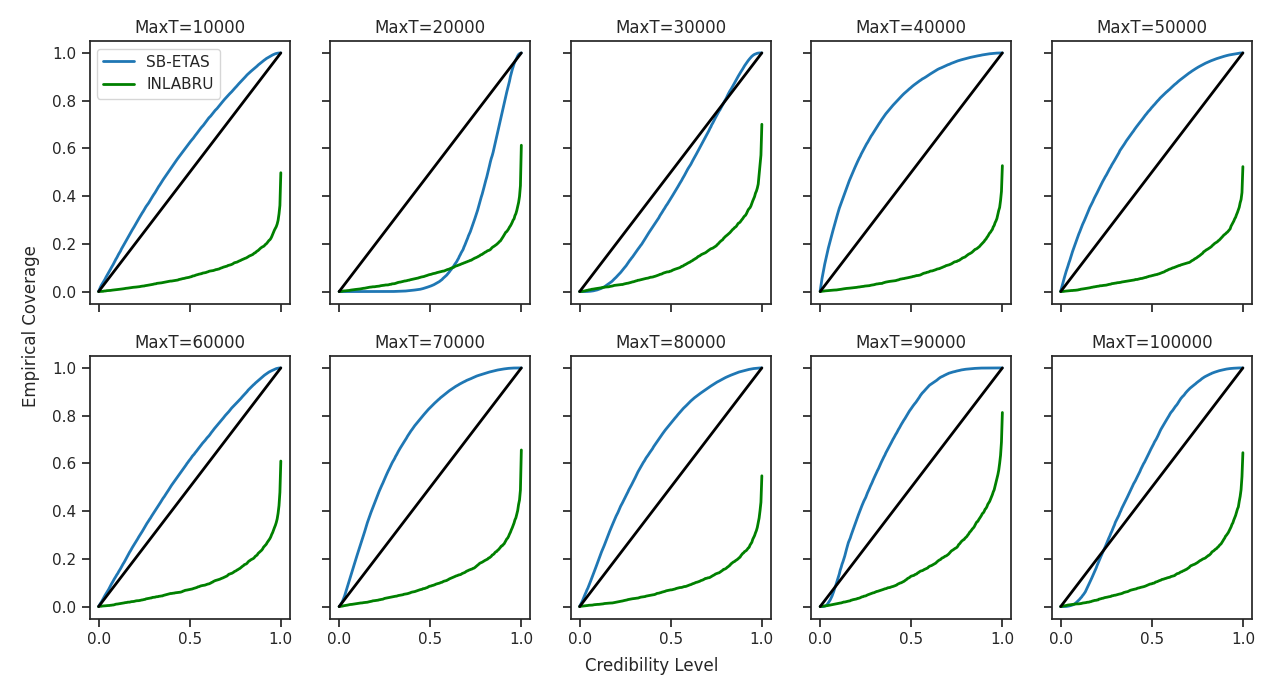}
    \caption{Empirical estimates of the coverage of both SB-ETAS and \texttt{inlabru}. Coverage below the black line $y=x$ indicates an overconfident approximation, whereas coverage below $y=x$ indicates a conservative approximation.}
    \label{fig:coverage}
\end{figure*}
A common measure for the appropriateness of a prediction's uncertainty is the coverage property \citep{prangle2014diagnostic,xing2019calibrated}. The coverage of an approximate posterior assesses the quality of its credible regions $\hat{C}_{\textbf{x}_{obs}}$ which satisfy,
\begin{equation}\label{eq:credibility}
    \gamma = \mathbb{E}_{\hat{p}(\theta|\textbf{x}_{obs})}\left(\mathbb{I}\{\hat{\theta}\in \hat{C}_{\textbf{x}_{obs}}\}\right) 
\end{equation}
An approximate posterior has perfect coverage if its operational coverage,
\begin{equation}\label{eq:operational_coverage}
    b(\textbf{x}_{obs}) = \mathbb{E}_{p(\theta|\textbf{x}_{obs})}\left(\mathbb{I}\{\hat{\theta}\in \hat{C}_{\textbf{x}_{obs}}\}\right) 
\end{equation}
is equal to the credibility level $\gamma$. The approximation is conservative if it has operational coverage $b(\textbf{x}_{obs}) > \gamma$ and is overconfident if $b(\textbf{x}_{obs}) < \gamma$ \citep{hermans2021trust}. Expectations in equations (\ref{eq:credibility})-(\ref{eq:operational_coverage}) cannot be computed exactly and so are replaced with Monte Carlo averages, resulting in empirical coverage $c(\textbf{x}_{obs})$. Figure \ref{fig:coverage} shows the empirical coverage for both SB-ETAS, averaged across the 10 initial seeds, along with \verb|inlabru| on the 10 synthetic catalogs. \verb|inlabru| consistently gives overconfident approximations, as the empirical coverage lies well below the credibility level. SB-ETAS has empirical coverage that indicates conservative estimates, but that is generally closer to the credibility level.

\section{SCEDC Catalog}\label{sec:SCEDC}
We now evaluate SB-ETAS on some observational data from Southern California. The Southern California Seismic Network has produced an earthquake catalog for Southern California going back to 1932 \citep{hutton2010earthquake}. This catalog contains many famous large earthquakes such as the 1992 $M_W$ 7.3 Landers, 1999
$M_W$ 7.1 Hector Mine and $M_W$7.1 Ridgecrest sequences. We use $N=43,537$ events from \date{01/01/1981} - \date{31/12/2021} with earthquake magnitudes $\geq M_W\ 2.5$ since this assures the most data completeness \citep{hutton2010earthquake}. The catalog can be downloaded from the Southern California Earthquake Data Center \url{https://service.scedc.caltech.edu/ftp/catalogs/SCSN/}.

This size of catalog contains too many events to find ETAS posteriors using \verb|bayesianETAS| (i.e. it would take longer than 2 weeks). Therefore we run only SB-ETAS and \verb|inlabru| on the entire catalog and validate their performance by comparing the compensator, $\Lambda^*(t;\theta) = \int_0^t \lambda^*(s;\theta)ds$, with the observed cumulative number of events in the catalog $N(t)$. $\Lambda^*(t;\theta)$ gives the expected number of events at time $t$, and therefore a model and its parameters are consistent with the observed data if $\Lambda^*(t;\theta) = N(t)$.

\begin{figure*}[ht!]
    \centering
    \includegraphics[width=\textwidth]{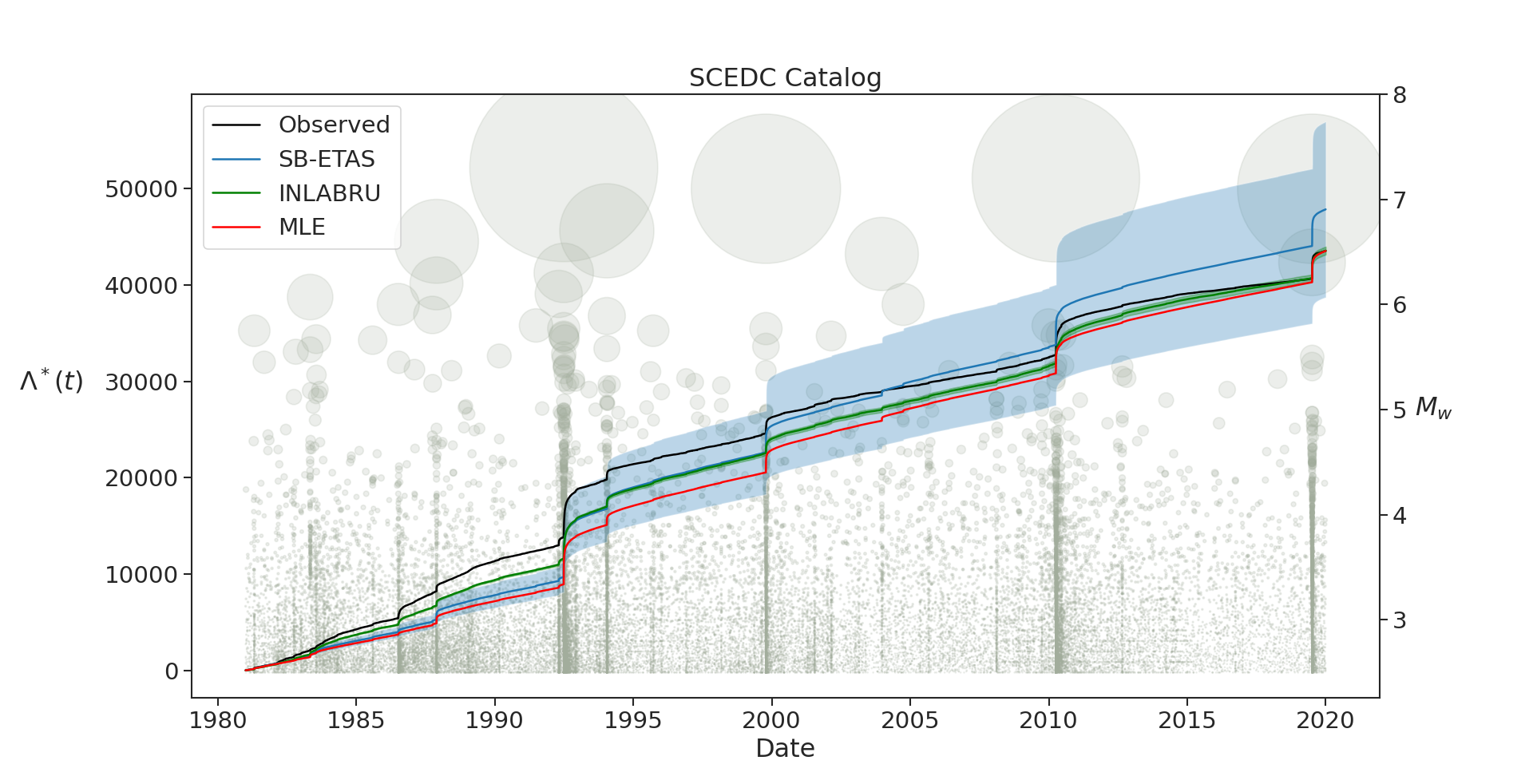}
    \caption{The compensator $\Lambda^*(t)$ found from estimating the ETAS posterior distribution on the SCEDC catalog (events displayed in background). 5,000 Samples from the posterior using both SB-ETAS and \texttt{inlabru} were used to generate a mean and 95\% confidence interval. The compensator is compared against the observed cumulative number of events in the catalog along with the MLE.}
    \label{fig:SCEDC_Lambdas}
\end{figure*}

We generate 5,000 samples using SB-ETAS and \verb|inlabru| and use each sample to generate a compensator curve $\Lambda^*(t;\theta)$. We display $95\%$ confidence intervals of these curves in Figure \ref{fig:SCEDC_Lambdas}, along with a curve for the Maximum Likelihood Estimate (MLE). Consistent with the synthetic experiments, we find that SB-ETAS gives a conservative estimate of the cumulative number of events across the catalog, whereas \verb|inlabru| is overconfident and does not contain the observed number of events within its very narrow confidence interval. Both \verb|inlabru| and the MLE match the total observed number of events in the catalog, since this value, $\Lambda^*(T)$, is a dominant term in each of their loss functions (the likelihood) during estimation.

\begin{figure}[h]
    \centering
    \includegraphics[width=0.5\textwidth]{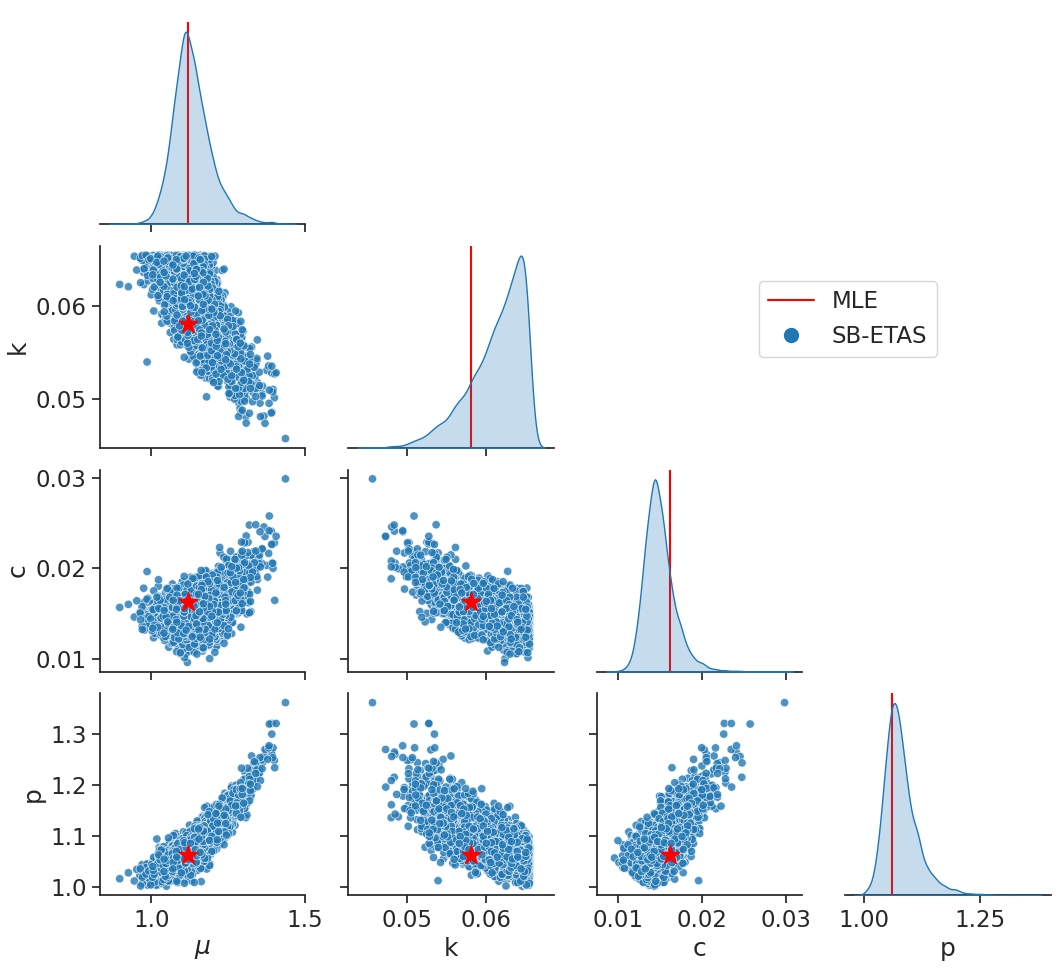}
    \caption{The posterior distribution of ETAS parameters found on the SCEDC catalog using SB-ETAS. This implementation of ETAS fixes $\alpha = \beta$. MLE parameters are plotted for comparison.}
    \label{fig:SB-ETAS SCEDC samples}
\end{figure}

\begin{figure}[h]
    \centering
    \includegraphics[width=0.5\textwidth]{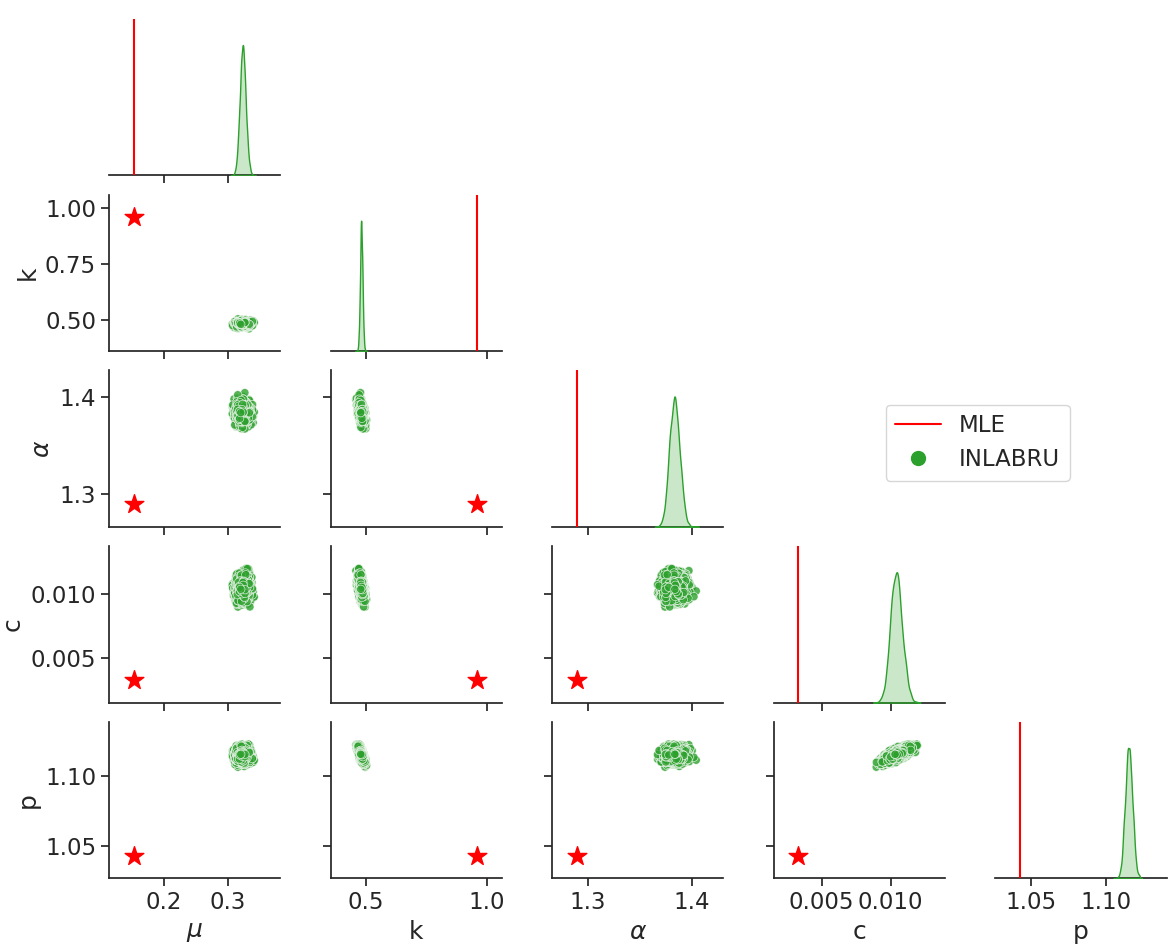}
    \caption{The posterior distribution of ETAS parameters found on the SCEDC catalog using \texttt{inlabru}. This implementation of ETAS has a free $\alpha$ parameter. MLE parameters are plotted for comparison.}
    \label{fig:INLABRU SCEDC samples}
\end{figure}

For both the MLE and SB-ETAS, we fix the $\alpha$ parameter equal to the $\beta$ parameter of the Gutenberg-Richter law $f_{GR(m)}$, a result that is consistent with other temporal only studies of Southern California \citep{felzer2004origin,helmstetter2005triggering}, and which reproduces Båth’s law for aftershocks \citep{felzer2002triggering}.
We were unable to successfully fix $\alpha$ for \verb|inlabru| and therefore use the 5 parameter implementation of ETAS. Posterior distributions are displayed in Figures \ref{fig:SB-ETAS SCEDC samples} and \ref{fig:INLABRU SCEDC samples} including $\alpha = \beta$ and free $\alpha$ implementations of the MLE. Although the modes of the marginal distributions do not match the MLE, the SB-ETAS posteriors contain the MLE parameters within their wider confidence ranges. Since \verb|inlabru| has much narrower confidence, although the distributions are relatively close to the MLE, the confidence ranges do not contain MLE parameters.


\section{Discussion and Conclusion}\label{sec:D&C}
The growing size of earthquake catalogs generated through machine learning based phase picking and an increased density of seismic networks has overtaken the ability of our current forecasting models to fit to such large datasets. Although improvements have been made to reduce the computational time of performing Bayesian inference for the ETAS model, first with \verb|bayesianETAS| followed by \verb|inlabru|, neither of these approaches improve upon the scalability of inference. Therefore as catalogs continue to grow in size, these methods become less feasible to use. On experiments where we give SB-ETAS, \verb|bayesianETAS| and \verb|inlabru| access to the same 8 CPUs, only SB-ETAS could be used to fit a catalog of 500,000 events and was the fastest method for catalogs above 100,000 events. Both \verb|inlabru| and SB-ETAS are parallelized methods and would therefore see a reduction in runtime if given access to more CPUs. This is unlike \verb|bayesianETAS| which is not parrallelized in its current implementation. It is also worth noting that although SB-ETAS and \verb|inlabru| were given the same CPUs, \verb|inlabru| required over 4 times the amount of memory than SB-ETAS with catalogs over 100,000 events, (Figure \ref{fig:memory}). This additional memory demand far exceeds the capacity typically available on standard laptops.


Comparing samples from both approximate methods to samples from \verb|bayesianETAS| demonstrates the quality of each inference method. Our general finding is that \verb|inlabru| provides overconfident estimates of posteriors whereas SB-ETAS gives conservative estimates. Although it might seem reasonable to judge an approximate posterior by its closeness to the exact posterior, for practical use, overconfident estimates should be penalised more than under-confident ones. Bayesian inference for the ETAS model seeks to identify a range of parameter values which are then used to give confidence over a range of forecasts. However, failure to identify regions of the parameter space that give likely parameters would result in omission of a range of likely forecasts.

SB-ETAS provides an efficient procedure for performing Bayesian Inference for a temporal ETAS model on large earthquake catalogs. However, operational earthquake forecasting typically requires spatial forecasts for target regions that are large enough to exhibit significant spatial variability. The spatio-temporal form of the ETAS model extends the temporal model used in the study by modeling earthquake spatial interactions with an isotropic Gaussian spatial triggering kernel \citep{ogata1998space}. This spatio-temporal ETAS model is also defined as a branching process and would therefore retain the $\mathcal{O}(n\log n)$ complexity of simulation. This study has illustrated that the Ripley K-statistic is an informative summary statistic for the triggering parameters of the temporal ETAS model. It seems fair to assume that the spatio-temporal Ripley K-statistic,
\begin{align*}
    &\hat{K}(\textbf{x},w_t,w_s) = \\ &\hspace{0.2cm}\frac{A T}{n^2}\sum_{i=1}^n \sum_{j\neq i}\mathbb{I}(0< t_j-t_i \leq w_t)\mathbb{I}(||s_j-s_i||_2 \leq w_s).
\end{align*}
where $A$ is the area of the study region, would be a reasonable choice for the spatio-temporal form of SB-ETAS. This statistic loses the $\mathcal{O}(n)$ efficiency that the purely temporal one benefits from. Instead \cite{wang2020optimizing} have developed a distributed procedure for calculating this statistic with $\mathcal{O}(n\log n)$ complexity that would retain the overall time complexity that SB-ETAS has.

Ideally, the value of the Ripley K-statistic $\hat{K}(\textbf{x},w)$ for all $w \in \mathbb{R}_+$ would be used as the summary statistic for the observed data $\textbf{x}$. However, since the neural density estimator requires a fixed length vector as input, we have to sample this function at pre-specified intervals. Increasing the number of samples would increase the dimension of this fixed length vector, making the density estimation task more challenging. On the other hand, using less samples $w$, would make the density estimation task easier but would reduce the information contained in the summary statistic. Future work, should look at balancing the number of samples of the Ripley K-statistic as well as moving beyond the hand chosen values used in this study. We speculate that the loss of information from under-sampling the K-statistic, weakens the generalisation of the method in its current form, e.g. the MMD for the MaxT=30000 experiment does not decreasing over the simulation rounds (Figure \ref{fig:mmd}). 

SB-ETAS demonstrates that inference for parameters of a clustered temporal point process can be achieved without having to evaluate an expensive likelihood function. However, generally simulation based inference is used in cases where the likelihood function is intractable. For example, consider the same ETAS branching process used in this study, but instead events are deleted with a time varying probability $h(t)$. The induced likelihood of this process,
\begin{align}\label{eq:intractable}
    p(\textbf{x}|\theta) \propto \int p(\textbf{x},\textbf{x}_u|\theta) \prod_{t_i \in \textbf{x}}h(t_i)\prod_{t_j \in \textbf{x}_u}(1-h(t_j))d\textbf{x}_u,
\end{align}
is intractable since it involves integrating over the set of unobserved events $\textbf{x}_u$ \citep{deutsch2021abc}. This time-varying detection of events is a source of bias in earthquake models and current methods to deal with it typically estimate the true earthquake rate from the apparent earthquake rate assuming no contribution from undetected events \citep{hainzl2016apparent}. There are biases from ignoring such triggering \citep{sornette2005apparent} which could be avoided by using this simulation based random deletion model with intractable likelihood Eq. \ref{eq:intractable}.

Another application of this simulation based framework is to earthquake branching models that include complex physical dependencies. One possible example would be to calibrate the Third Uniform California Earthquake Rupture Forecast ETAS Model (UCERF3-ETAS), a unified model for fault rupture and ETAS earthquake clustering \citep{field2017spatiotemporal}. This model extends the standard ETAS model by explicitly modeling fault ruptures in California and includes a variable magnitude distribution which significantly affects the triggering probabilities of large earthquakes. This model is only defined as a simulator and uses ETAS parameters found independently to the joint ETAS and fault model.  In fact, \cite{page2018turing} validate the models performance through a comparison of summary statistics from the outputs of the model. This validation could be extended to comprise part of the inference procedure for model parameters using the same simulation based framework as SB-ETAS.

\begin{appendices}

\section{Memory}\label{secA1}

\begin{figure}[h!]
    \centering
    \includegraphics[width=0.5\textwidth]{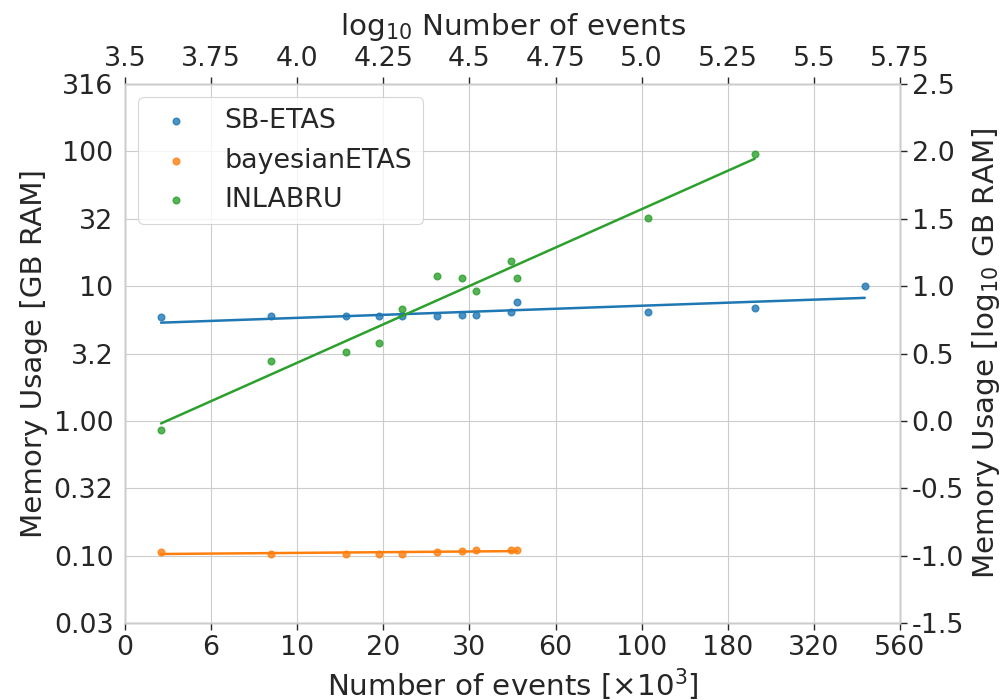}
    \caption{The memory usage for parameter inference versus the catalog size for SB-ETAS, \texttt{inlabru} and \texttt{bayesianETAS}. Separate ETAS catalogs were generated with the same intensity function parameters but for varying size time-windows. The memory usage and the number of events are plotted in log-log space.}
    \label{fig:memory}
\end{figure}




\end{appendices}


\bibliography{sn-bibliography}


\begin{thebibliography}{76}
\ifx \bisbn   \undefined \def \bisbn  #1{ISBN #1}\fi
\ifx \binits  \undefined \def \binits#1{#1}\fi
\ifx \bauthor  \undefined \def \bauthor#1{#1}\fi
\ifx \batitle  \undefined \def \batitle#1{#1}\fi
\ifx \bjtitle  \undefined \def \bjtitle#1{#1}\fi
\ifx \bvolume  \undefined \def \bvolume#1{\textbf{#1}}\fi
\ifx \byear  \undefined \def \byear#1{#1}\fi
\ifx \bissue  \undefined \def \bissue#1{#1}\fi
\ifx \bfpage  \undefined \def \bfpage#1{#1}\fi
\ifx \blpage  \undefined \def \blpage #1{#1}\fi
\ifx \burl  \undefined \def \burl#1{\textsf{#1}}\fi
\ifx \doiurl  \undefined \def \doiurl#1{\url{https://doi.org/#1}}\fi
\ifx \betal  \undefined \def \betal{\textit{et al.}}\fi
\ifx \binstitute  \undefined \def \binstitute#1{#1}\fi
\ifx \binstitutionaled  \undefined \def \binstitutionaled#1{#1}\fi
\ifx \bctitle  \undefined \def \bctitle#1{#1}\fi
\ifx \beditor  \undefined \def \beditor#1{#1}\fi
\ifx \bpublisher  \undefined \def \bpublisher#1{#1}\fi
\ifx \bbtitle  \undefined \def \bbtitle#1{#1}\fi
\ifx \bedition  \undefined \def \bedition#1{#1}\fi
\ifx \bseriesno  \undefined \def \bseriesno#1{#1}\fi
\ifx \blocation  \undefined \def \blocation#1{#1}\fi
\ifx \bsertitle  \undefined \def \bsertitle#1{#1}\fi
\ifx \bsnm \undefined \def \bsnm#1{#1}\fi
\ifx \bsuffix \undefined \def \bsuffix#1{#1}\fi
\ifx \bparticle \undefined \def \bparticle#1{#1}\fi
\ifx \barticle \undefined \def \barticle#1{#1}\fi
\bibcommenthead
\ifx \bconfdate \undefined \def \bconfdate #1{#1}\fi
\ifx \botherref \undefined \def \botherref #1{#1}\fi
\ifx \url \undefined \def \url#1{\textsf{#1}}\fi
\ifx \bchapter \undefined \def \bchapter#1{#1}\fi
\ifx \bbook \undefined \def \bbook#1{#1}\fi
\ifx \bcomment \undefined \def \bcomment#1{#1}\fi
\ifx \oauthor \undefined \def \oauthor#1{#1}\fi
\ifx \citeauthoryear \undefined \def \citeauthoryear#1{#1}\fi
\ifx \endbibitem  \undefined \def \endbibitem {}\fi
\ifx \bconflocation  \undefined \def \bconflocation#1{#1}\fi
\ifx \arxivurl  \undefined \def \arxivurl#1{\textsf{#1}}\fi
\csname PreBibitemsHook\endcsname

\bibitem[\protect\citeauthoryear{Bacry et~al.}{2016}]{bacry2016estimation}
\begin{barticle}
\bauthor{\bsnm{Bacry}, \binits{E.}},
\bauthor{\bsnm{Jaisson}, \binits{T.}},
\bauthor{\bsnm{Muzy}, \binits{J.-F.}}:
\batitle{Estimation of slowly decreasing hawkes kernels: application to high-frequency order book dynamics}.
\bjtitle{Quantitative Finance}
\bvolume{16}(\bissue{8}),
\bfpage{1179}--\blpage{1201}
(\byear{2016})
\end{barticle}
\endbibitem

\bibitem[\protect\citeauthoryear{Bacry and Muzy}{2014}]{bacry2014second}
\begin{botherref}
\oauthor{\bsnm{Bacry}, \binits{E.}},
\oauthor{\bsnm{Muzy}, \binits{J.-F.}}:
Second order statistics characterization of hawkes processes and non-parametric estimation.
arXiv preprint arXiv:1401.0903
(2014)
\end{botherref}
\endbibitem

\bibitem[\protect\citeauthoryear{Beaumont et~al.}{2002}]{beaumont2002approximate}
\begin{barticle}
\bauthor{\bsnm{Beaumont}, \binits{M.A.}},
\bauthor{\bsnm{Zhang}, \binits{W.}},
\bauthor{\bsnm{Balding}, \binits{D.J.}}:
\batitle{Approximate bayesian computation in population genetics}.
\bjtitle{Genetics}
\bvolume{162}(\bissue{4}),
\bfpage{2025}--\blpage{2035}
(\byear{2002})
\end{barticle}
\endbibitem

\bibitem[\protect\citeauthoryear{Cranmer et~al.}{2020}]{cranmer2020frontier}
\begin{barticle}
\bauthor{\bsnm{Cranmer}, \binits{K.}},
\bauthor{\bsnm{Brehmer}, \binits{J.}},
\bauthor{\bsnm{Louppe}, \binits{G.}}:
\batitle{The frontier of simulation-based inference}.
\bjtitle{Proceedings of the National Academy of Sciences}
\bvolume{117}(\bissue{48}),
\bfpage{30055}--\blpage{30062}
(\byear{2020})
\end{barticle}
\endbibitem

\bibitem[\protect\citeauthoryear{Cattania et~al.}{2018}]{cattania2018forecasting}
\begin{barticle}
\bauthor{\bsnm{Cattania}, \binits{C.}},
\bauthor{\bsnm{Werner}, \binits{M.J.}},
\bauthor{\bsnm{Marzocchi}, \binits{W.}},
\bauthor{\bsnm{Hainzl}, \binits{S.}},
\bauthor{\bsnm{Rhoades}, \binits{D.}},
\bauthor{\bsnm{Gerstenberger}, \binits{M.}},
\bauthor{\bsnm{Liukis}, \binits{M.}},
\bauthor{\bsnm{Savran}, \binits{W.}},
\bauthor{\bsnm{Christophersen}, \binits{A.}},
\bauthor{\bsnm{Helmstetter}, \binits{A.}}, \betal:
\batitle{The forecasting skill of physics-based seismicity models during the 2010--2012 canterbury, new zealand, earthquake sequence}.
\bjtitle{Seismological Research Letters}
\bvolume{89}(\bissue{4}),
\bfpage{1238}--\blpage{1250}
(\byear{2018})
\end{barticle}
\endbibitem

\bibitem[\protect\citeauthoryear{Diggle}{1985}]{diggle1985kernel}
\begin{barticle}
\bauthor{\bsnm{Diggle}, \binits{P.}}:
\batitle{A kernel method for smoothing point process data}.
\bjtitle{Journal of the Royal Statistical Society: Series C (Applied Statistics)}
\bvolume{34}(\bissue{2}),
\bfpage{138}--\blpage{147}
(\byear{1985})
\end{barticle}
\endbibitem

\bibitem[\protect\citeauthoryear{Dixon}{2001}]{dixon2001ripley}
\begin{botherref}
\oauthor{\bsnm{Dixon}, \binits{P.}}:
Ripley’s k function
(2001)
\end{botherref}
\endbibitem

\bibitem[\protect\citeauthoryear{Deutsch and Ross}{2021}]{deutsch2021abc}
\begin{botherref}
\oauthor{\bsnm{Deutsch}, \binits{I.}},
\oauthor{\bsnm{Ross}, \binits{G.J.}}:
ABC Learning of Hawkes Processes with Missing or Noisy Event Times
(2021)
\end{botherref}
\endbibitem

\bibitem[\protect\citeauthoryear{Ertekin et~al.}{2015}]{ertekin2015reactive}
\begin{botherref}
\oauthor{\bsnm{Ertekin}, \binits{{\c{S}}.}},
\oauthor{\bsnm{Rudin}, \binits{C.}},
\oauthor{\bsnm{McCormick}, \binits{T.H.}}:
Reactive point processes: A new approach to predicting power failures in underground electrical systems
(2015)
\end{botherref}
\endbibitem

\bibitem[\protect\citeauthoryear{Felzer et~al.}{2004}]{felzer2004origin}
\begin{barticle}
\bauthor{\bsnm{Felzer}, \binits{K.}},
\bauthor{\bsnm{Abercrombie}, \binits{R.}},
\bauthor{\bsnm{Ekstrom}, \binits{G.}}:
\batitle{A common origin for aftershocks, foreshocks, and multiplets}.
\bjtitle{Bulletin of The Seismological Society of America - BULL SEISMOL SOC AMER}
\bvolume{94},
\bfpage{88}--\blpage{98}
(\byear{2004})
\doiurl{10.1785/0120030069}
\end{barticle}
\endbibitem

\bibitem[\protect\citeauthoryear{Felzer et~al.}{2002}]{felzer2002triggering}
\begin{barticle}
\bauthor{\bsnm{Felzer}, \binits{K.R.}},
\bauthor{\bsnm{Becker}, \binits{T.W.}},
\bauthor{\bsnm{Abercrombie}, \binits{R.E.}},
\bauthor{\bsnm{Ekstr{\"o}m}, \binits{G.}},
\bauthor{\bsnm{Rice}, \binits{J.R.}}:
\batitle{Triggering of the 1999 mw 7.1 hector mine earthquake by aftershocks of the 1992 mw 7.3 landers earthquake}.
\bjtitle{Journal of Geophysical Research: Solid Earth}
\bvolume{107}(\bissue{B9}),
\bfpage{6}
(\byear{2002})
\end{barticle}
\endbibitem

\bibitem[\protect\citeauthoryear{Field et~al.}{2017}]{field2017synoptic}
\begin{barticle}
\bauthor{\bsnm{Field}, \binits{E.H.}},
\bauthor{\bsnm{Jordan}, \binits{T.H.}},
\bauthor{\bsnm{Page}, \binits{M.T.}},
\bauthor{\bsnm{Milner}, \binits{K.R.}},
\bauthor{\bsnm{Shaw}, \binits{B.E.}},
\bauthor{\bsnm{Dawson}, \binits{T.E.}},
\bauthor{\bsnm{Biasi}, \binits{G.P.}},
\bauthor{\bsnm{Parsons}, \binits{T.}},
\bauthor{\bsnm{Hardebeck}, \binits{J.L.}},
\bauthor{\bsnm{Michael}, \binits{A.J.}}, \betal:
\batitle{A synoptic view of the third uniform california earthquake rupture forecast (ucerf3)}.
\bjtitle{Seismological Research Letters}
\bvolume{88}(\bissue{5}),
\bfpage{1259}--\blpage{1267}
(\byear{2017})
\end{barticle}
\endbibitem

\bibitem[\protect\citeauthoryear{Field et~al.}{2017}]{field2017spatiotemporal}
\begin{barticle}
\bauthor{\bsnm{Field}, \binits{E.H.}},
\bauthor{\bsnm{Milner}, \binits{K.R.}},
\bauthor{\bsnm{Hardebeck}, \binits{J.L.}},
\bauthor{\bsnm{Page}, \binits{M.T.}},
\bauthor{\bsnm{Elst}, \binits{N.}},
\bauthor{\bsnm{Jordan}, \binits{T.H.}},
\bauthor{\bsnm{Michael}, \binits{A.J.}},
\bauthor{\bsnm{Shaw}, \binits{B.E.}},
\bauthor{\bsnm{Werner}, \binits{M.J.}}:
\batitle{A spatiotemporal clustering model for the third uniform california earthquake rupture forecast (ucerf3-etas): Toward an operational earthquake forecast}.
\bjtitle{Bulletin of the Seismological Society of America}
\bvolume{107}(\bissue{3}),
\bfpage{1049}--\blpage{1081}
(\byear{2017})
\end{barticle}
\endbibitem

\bibitem[\protect\citeauthoryear{Gretton et~al.}{2012}]{gretton2012kernel}
\begin{barticle}
\bauthor{\bsnm{Gretton}, \binits{A.}},
\bauthor{\bsnm{Borgwardt}, \binits{K.M.}},
\bauthor{\bsnm{Rasch}, \binits{M.J.}},
\bauthor{\bsnm{Sch{\"o}lkopf}, \binits{B.}},
\bauthor{\bsnm{Smola}, \binits{A.}}:
\batitle{A kernel two-sample test}.
\bjtitle{The Journal of Machine Learning Research}
\bvolume{13}(\bissue{1}),
\bfpage{723}--\blpage{773}
(\byear{2012})
\end{barticle}
\endbibitem

\bibitem[\protect\citeauthoryear{Geffner et~al.}{2022}]{geffner2022score}
\begin{bchapter}
\bauthor{\bsnm{Geffner}, \binits{T.}},
\bauthor{\bsnm{Papamakarios}, \binits{G.}},
\bauthor{\bsnm{Mnih}, \binits{A.}}:
\bctitle{Score modeling for simulation-based inference}.
In: \bbtitle{NeurIPS 2022 Workshop on Score-Based Methods}
(\byear{2022})
\end{bchapter}
\endbibitem

\bibitem[\protect\citeauthoryear{Gutenberg and Richter}{1936}]{gutenberg1936magnitude}
\begin{barticle}
\bauthor{\bsnm{Gutenberg}, \binits{B.}},
\bauthor{\bsnm{Richter}, \binits{C.F.}}:
\batitle{Magnitude and energy of earthquakes}.
\bjtitle{Science}
\bvolume{83}(\bissue{2147}),
\bfpage{183}--\blpage{185}
(\byear{1936})
\end{barticle}
\endbibitem

\bibitem[\protect\citeauthoryear{Hainzl}{2016a}]{hainzl2016apparent}
\begin{barticle}
\bauthor{\bsnm{Hainzl}, \binits{S.}}:
\batitle{Apparent triggering function of aftershocks resulting from rate-dependent incompleteness of earthquake catalogs}.
\bjtitle{Journal of Geophysical Research: Solid Earth}
\bvolume{121}(\bissue{9}),
\bfpage{6499}--\blpage{6509}
(\byear{2016})
\end{barticle}
\endbibitem

\bibitem[\protect\citeauthoryear{Hainzl}{2016b}]{hainzl2016rate}
\begin{barticle}
\bauthor{\bsnm{Hainzl}, \binits{S.}}:
\batitle{Rate-dependent incompleteness of earthquake catalogs}.
\bjtitle{Seismological Research Letters}
\bvolume{87}(\bissue{2A}),
\bfpage{337}--\blpage{344}
(\byear{2016})
\end{barticle}
\endbibitem

\bibitem[\protect\citeauthoryear{Hawkes}{1971}]{hawkes1971spectra}
\begin{barticle}
\bauthor{\bsnm{Hawkes}, \binits{A.G.}}:
\batitle{Spectra of some self-exciting and mutually exciting point processes}.
\bjtitle{Biometrika}
\bvolume{58}(\bissue{1}),
\bfpage{83}--\blpage{90}
(\byear{1971})
\end{barticle}
\endbibitem

\bibitem[\protect\citeauthoryear{Hermans et~al.}{2021}]{hermans2021trust}
\begin{botherref}
\oauthor{\bsnm{Hermans}, \binits{J.}},
\oauthor{\bsnm{Delaunoy}, \binits{A.}},
\oauthor{\bsnm{Rozet}, \binits{F.}},
\oauthor{\bsnm{Wehenkel}, \binits{A.}},
\oauthor{\bsnm{Begy}, \binits{V.}},
\oauthor{\bsnm{Louppe}, \binits{G.}}:
A trust crisis in simulation-based inference? your posterior approximations can be unfaithful.
arXiv preprint arXiv:2110.06581
(2021)
\end{botherref}
\endbibitem

\bibitem[\protect\citeauthoryear{Helmstetter et~al.}{2005}]{helmstetter2005triggering}
\begin{botherref}
\oauthor{\bsnm{Helmstetter}, \binits{A.}},
\oauthor{\bsnm{Kagan}, \binits{Y.Y.}},
\oauthor{\bsnm{Jackson}, \binits{D.D.}}:
Importance of small earthquakes for stress transfers and earthquake triggering.
Journal of Geophysical Research: Solid Earth
\textbf{110}(B5)
(2005)
\doiurl{10.1029/2004JB003286}
{\href{https://arxiv.org/abs/https://agupubs.onlinelibrary.wiley.com/doi/pdf/10.1029/2004JB003286}{{https://agupubs.onlinelibrary.wiley.com/doi/pdf/10.1029/2004JB003286}}}
\end{botherref}
\endbibitem

\bibitem[\protect\citeauthoryear{Hutton et~al.}{2010}]{hutton2010earthquake}
\begin{barticle}
\bauthor{\bsnm{Hutton}, \binits{K.}},
\bauthor{\bsnm{Woessner}, \binits{J.}},
\bauthor{\bsnm{Hauksson}, \binits{E.}}:
\batitle{Earthquake monitoring in southern california for seventy-seven years (1932--2008)}.
\bjtitle{Bulletin of the Seismological Society of America}
\bvolume{100}(\bissue{2}),
\bfpage{423}--\blpage{446}
(\byear{2010})
\end{barticle}
\endbibitem

\bibitem[\protect\citeauthoryear{Iturrieta et~al.}{2024}]{iturrieta2024evaluation}
\begin{botherref}
\oauthor{\bsnm{Iturrieta}, \binits{P.}},
\oauthor{\bsnm{Bayona}, \binits{J.A.}},
\oauthor{\bsnm{Werner}, \binits{M.J.}},
\oauthor{\bsnm{Schorlemmer}, \binits{D.}},
\oauthor{\bsnm{Taroni}, \binits{M.}},
\oauthor{\bsnm{Falcone}, \binits{G.}},
\oauthor{\bsnm{Cotton}, \binits{F.}},
\oauthor{\bsnm{Khawaja}, \binits{A.M.}},
\oauthor{\bsnm{Savran}, \binits{W.H.}},
\oauthor{\bsnm{Marzocchi}, \binits{W.}}:
Evaluation of a decade-long prospective earthquake forecasting experiment in italy.
Seismological Research Letters
(2024)
\end{botherref}
\endbibitem

\bibitem[\protect\citeauthoryear{Ide}{2013}]{ide2013proportionality}
\begin{barticle}
\bauthor{\bsnm{Ide}, \binits{S.}}:
\batitle{The proportionality between relative plate velocity and seismicity in subduction zones}.
\bjtitle{Nature Geoscience}
\bvolume{6}(\bissue{9}),
\bfpage{780}--\blpage{784}
(\byear{2013})
\end{barticle}
\endbibitem

\bibitem[\protect\citeauthoryear{Izbicki et~al.}{2014}]{izbicki2014high}
\begin{bchapter}
\bauthor{\bsnm{Izbicki}, \binits{R.}},
\bauthor{\bsnm{Lee}, \binits{A.}},
\bauthor{\bsnm{Schafer}, \binits{C.}}:
\bctitle{High-dimensional density ratio estimation with extensions to approximate likelihood computation}.
In: \bbtitle{Artificial Intelligence and Statistics},
pp. \bfpage{420}--\blpage{429}
(\byear{2014}).
\bcomment{PMLR}
\end{bchapter}
\endbibitem

\bibitem[\protect\citeauthoryear{Lueckmann et~al.}{2021}]{lueckmann2021benchmarking}
\begin{bchapter}
\bauthor{\bsnm{Lueckmann}, \binits{J.-M.}},
\bauthor{\bsnm{Boelts}, \binits{J.}},
\bauthor{\bsnm{Greenberg}, \binits{D.}},
\bauthor{\bsnm{Goncalves}, \binits{P.}},
\bauthor{\bsnm{Macke}, \binits{J.}}:
\bctitle{Benchmarking simulation-based inference}.
In: \bbtitle{International Conference on Artificial Intelligence and Statistics},
pp. \bfpage{343}--\blpage{351}
(\byear{2021}).
\bcomment{PMLR}
\end{bchapter}
\endbibitem

\bibitem[\protect\citeauthoryear{Lueckmann et~al.}{2017}]{lueckmann2017flexible}
\begin{botherref}
\oauthor{\bsnm{Lueckmann}, \binits{J.-M.}},
\oauthor{\bsnm{Goncalves}, \binits{P.J.}},
\oauthor{\bsnm{Bassetto}, \binits{G.}},
\oauthor{\bsnm{{\"O}cal}, \binits{K.}},
\oauthor{\bsnm{Nonnenmacher}, \binits{M.}},
\oauthor{\bsnm{Macke}, \binits{J.H.}}:
Flexible statistical inference for mechanistic models of neural dynamics.
Advances in neural information processing systems
\textbf{30}
(2017)
\end{botherref}
\endbibitem

\bibitem[\protect\citeauthoryear{Lombardi}{2015}]{lombardi2015estimation}
\begin{barticle}
\bauthor{\bsnm{Lombardi}, \binits{A.M.}}:
\batitle{Estimation of the parameters of etas models by simulated annealing}.
\bjtitle{Scientific reports}
\bvolume{5}(\bissue{1}),
\bfpage{8417}
(\byear{2015})
\end{barticle}
\endbibitem

\bibitem[\protect\citeauthoryear{Lopez-Paz and Oquab}{2016}]{lopez2016revisiting}
\begin{botherref}
\oauthor{\bsnm{Lopez-Paz}, \binits{D.}},
\oauthor{\bsnm{Oquab}, \binits{M.}}:
Revisiting classifier two-sample tests.
arXiv preprint arXiv:1610.06545
(2016)
\end{botherref}
\endbibitem

\bibitem[\protect\citeauthoryear{Lehmann and Romano}{2005}]{lehmann2005testing}
\begin{bbook}
\bauthor{\bsnm{Lehmann}, \binits{E.L.}},
\bauthor{\bsnm{Romano}, \binits{J.P.}}:
\bbtitle{Testing Statistical Hypotheses},
\bedition{3}rd edn.
\bsertitle{Springer Texts in Statistics},
p. \bfpage{784}.
\bpublisher{Springer},
\blocation{New York}
(\byear{2005})
\end{bbook}
\endbibitem

\bibitem[\protect\citeauthoryear{Molkenthin et~al.}{2022}]{molkenthin2022gp}
\begin{barticle}
\bauthor{\bsnm{Molkenthin}, \binits{C.}},
\bauthor{\bsnm{Donner}, \binits{C.}},
\bauthor{\bsnm{Reich}, \binits{S.}},
\bauthor{\bsnm{Z{\"o}ller}, \binits{G.}},
\bauthor{\bsnm{Hainzl}, \binits{S.}},
\bauthor{\bsnm{Holschneider}, \binits{M.}},
\bauthor{\bsnm{Opper}, \binits{M.}}:
\batitle{Gp-etas: semiparametric bayesian inference for the spatio-temporal epidemic type aftershock sequence model}.
\bjtitle{Statistics and computing}
\bvolume{32}(\bissue{2}),
\bfpage{29}
(\byear{2022})
\end{barticle}
\endbibitem

\bibitem[\protect\citeauthoryear{Marzocchi et~al.}{2014}]{marzocchi2014establishment}
\begin{barticle}
\bauthor{\bsnm{Marzocchi}, \binits{W.}},
\bauthor{\bsnm{Lombardi}, \binits{A.M.}},
\bauthor{\bsnm{Casarotti}, \binits{E.}}:
\batitle{The establishment of an operational earthquake forecasting system in italy}.
\bjtitle{Seismological Research Letters}
\bvolume{85}(\bissue{5}),
\bfpage{961}--\blpage{969}
(\byear{2014})
\end{barticle}
\endbibitem

\bibitem[\protect\citeauthoryear{Marjoram et~al.}{2003}]{marjoram2003markov}
\begin{barticle}
\bauthor{\bsnm{Marjoram}, \binits{P.}},
\bauthor{\bsnm{Molitor}, \binits{J.}},
\bauthor{\bsnm{Plagnol}, \binits{V.}},
\bauthor{\bsnm{Tavar{\'e}}, \binits{S.}}:
\batitle{Markov chain monte carlo without likelihoods}.
\bjtitle{Proceedings of the National Academy of Sciences}
\bvolume{100}(\bissue{26}),
\bfpage{15324}--\blpage{15328}
(\byear{2003})
\end{barticle}
\endbibitem

\bibitem[\protect\citeauthoryear{Mancini et~al.}{2019}]{mancini2019improving}
\begin{barticle}
\bauthor{\bsnm{Mancini}, \binits{S.}},
\bauthor{\bsnm{Segou}, \binits{M.}},
\bauthor{\bsnm{Werner}, \binits{M.}},
\bauthor{\bsnm{Cattania}, \binits{C.}}:
\batitle{Improving physics-based aftershock forecasts during the 2016--2017 central italy earthquake cascade}.
\bjtitle{Journal of Geophysical Research: Solid Earth}
\bvolume{124}(\bissue{8}),
\bfpage{8626}--\blpage{8643}
(\byear{2019})
\end{barticle}
\endbibitem

\bibitem[\protect\citeauthoryear{Mancini et~al.}{2020}]{mancini2020predictive}
\begin{barticle}
\bauthor{\bsnm{Mancini}, \binits{S.}},
\bauthor{\bsnm{Segou}, \binits{M.}},
\bauthor{\bsnm{Werner}, \binits{M.J.}},
\bauthor{\bsnm{Parsons}, \binits{T.}}:
\batitle{The predictive skills of elastic coulomb rate-and-state aftershock forecasts during the 2019 ridgecrest, california, earthquake sequence}.
\bjtitle{Bulletin of the Seismological Society of America}
\bvolume{110}(\bissue{4}),
\bfpage{1736}--\blpage{1751}
(\byear{2020})
\end{barticle}
\endbibitem

\bibitem[\protect\citeauthoryear{M{\o}ller and Waagepetersen}{2003}]{moller2003introduction}
\begin{bchapter}
\bauthor{\bsnm{M{\o}ller}, \binits{J.}},
\bauthor{\bsnm{Waagepetersen}, \binits{R.P.}}:
\bctitle{An introduction to simulation-based inference for spatial point processes}.
In: \bbtitle{Spatial Statistics and Computational Methods},
pp. \bfpage{143}--\blpage{198}.
\bpublisher{Springer}, \blocation{???}
(\byear{2003})
\end{bchapter}
\endbibitem

\bibitem[\protect\citeauthoryear{Ogata}{1978}]{ogata1978estimators}
\begin{barticle}
\bauthor{\bsnm{Ogata}, \binits{Y.}}:
\batitle{Estimators for stationary point processes}.
\bjtitle{Ann. Inst. Statist. Math}
\bvolume{30}(\bissue{Part A}),
\bfpage{243}--\blpage{261}
(\byear{1978})
\end{barticle}
\endbibitem

\bibitem[\protect\citeauthoryear{Ogata}{1988}]{ogata1988statistical}
\begin{barticle}
\bauthor{\bsnm{Ogata}, \binits{Y.}}:
\batitle{Statistical models for earthquake occurrences and residual analysis for point processes}.
\bjtitle{Journal of the American Statistical association}
\bvolume{83}(\bissue{401}),
\bfpage{9}--\blpage{27}
(\byear{1988})
\end{barticle}
\endbibitem

\bibitem[\protect\citeauthoryear{Ogata}{1998}]{ogata1998space}
\begin{barticle}
\bauthor{\bsnm{Ogata}, \binits{Y.}}:
\batitle{Space-time point-process models for earthquake occurrences}.
\bjtitle{Annals of the Institute of Statistical Mathematics}
\bvolume{50}(\bissue{2}),
\bfpage{379}--\blpage{402}
(\byear{1998})
\end{barticle}
\endbibitem

\bibitem[\protect\citeauthoryear{Omi et~al.}{2015}]{omi2015intermediate}
\begin{barticle}
\bauthor{\bsnm{Omi}, \binits{T.}},
\bauthor{\bsnm{Ogata}, \binits{Y.}},
\bauthor{\bsnm{Hirata}, \binits{Y.}},
\bauthor{\bsnm{Aihara}, \binits{K.}}:
\batitle{Intermediate-term forecasting of aftershocks from an early aftershock sequence: Bayesian and ensemble forecasting approaches}.
\bjtitle{Journal of Geophysical Research: Solid Earth}
\bvolume{120}(\bissue{4}),
\bfpage{2561}--\blpage{2578}
(\byear{2015})
\end{barticle}
\endbibitem

\bibitem[\protect\citeauthoryear{Omi et~al.}{2019}]{omi2019implementation}
\begin{barticle}
\bauthor{\bsnm{Omi}, \binits{T.}},
\bauthor{\bsnm{Ogata}, \binits{Y.}},
\bauthor{\bsnm{Shiomi}, \binits{K.}},
\bauthor{\bsnm{Enescu}, \binits{B.}},
\bauthor{\bsnm{Sawazaki}, \binits{K.}},
\bauthor{\bsnm{Aihara}, \binits{K.}}:
\batitle{Implementation of a real-time system for automatic aftershock forecasting in japan}.
\bjtitle{Seismological Research Letters}
\bvolume{90}(\bissue{1}),
\bfpage{242}--\blpage{250}
(\byear{2019})
\end{barticle}
\endbibitem

\bibitem[\protect\citeauthoryear{Prangle et~al.}{2014}]{prangle2014diagnostic}
\begin{barticle}
\bauthor{\bsnm{Prangle}, \binits{D.}},
\bauthor{\bsnm{Blum}, \binits{M.G.}},
\bauthor{\bsnm{Popovic}, \binits{G.}},
\bauthor{\bsnm{Sisson}, \binits{S.}}:
\batitle{Diagnostic tools for approximate bayesian computation using the coverage property}.
\bjtitle{Australian \& New Zealand Journal of Statistics}
\bvolume{56}(\bissue{4}),
\bfpage{309}--\blpage{329}
(\byear{2014})
\end{barticle}
\endbibitem

\bibitem[\protect\citeauthoryear{Prangle et~al.}{2014}]{prangle2014semi}
\begin{barticle}
\bauthor{\bsnm{Prangle}, \binits{D.}},
\bauthor{\bsnm{Fearnhead}, \binits{P.}},
\bauthor{\bsnm{Cox}, \binits{M.P.}},
\bauthor{\bsnm{Biggs}, \binits{P.J.}},
\bauthor{\bsnm{French}, \binits{N.P.}}:
\batitle{Semi-automatic selection of summary statistics for abc model choice}.
\bjtitle{Statistical applications in genetics and molecular biology}
\bvolume{13}(\bissue{1}),
\bfpage{67}--\blpage{82}
(\byear{2014})
\end{barticle}
\endbibitem

\bibitem[\protect\citeauthoryear{Papamakarios and Murray}{2016}]{papamakarios2016fast}
\begin{botherref}
\oauthor{\bsnm{Papamakarios}, \binits{G.}},
\oauthor{\bsnm{Murray}, \binits{I.}}:
Fast $\varepsilon$-free inference of simulation models with bayesian conditional density estimation.
Advances in neural information processing systems
\textbf{29}
(2016)
\end{botherref}
\endbibitem

\bibitem[\protect\citeauthoryear{Papamakarios et~al.}{2017}]{papamakarios2017masked}
\begin{botherref}
\oauthor{\bsnm{Papamakarios}, \binits{G.}},
\oauthor{\bsnm{Pavlakou}, \binits{T.}},
\oauthor{\bsnm{Murray}, \binits{I.}}:
Masked autoregressive flow for density estimation.
Advances in neural information processing systems
\textbf{30}
(2017)
\end{botherref}
\endbibitem

\bibitem[\protect\citeauthoryear{Papamakarios et~al.}{2019}]{papamakarios2019sequential}
\begin{bchapter}
\bauthor{\bsnm{Papamakarios}, \binits{G.}},
\bauthor{\bsnm{Sterratt}, \binits{D.}},
\bauthor{\bsnm{Murray}, \binits{I.}}:
\bctitle{Sequential neural likelihood: Fast likelihood-free inference with autoregressive flows}.
In: \bbtitle{The 22nd International Conference on Artificial Intelligence and Statistics},
pp. \bfpage{837}--\blpage{848}
(\byear{2019}).
\bcomment{PMLR}
\end{bchapter}
\endbibitem

\bibitem[\protect\citeauthoryear{Page and van~der Elst}{2018}]{page2018turing}
\begin{barticle}
\bauthor{\bsnm{Page}, \binits{M.T.}},
\bauthor{\bsnm{Elst}, \binits{N.J.}}:
\batitle{Turing-style tests for ucerf3 synthetic catalogs}.
\bjtitle{Bulletin of the Seismological Society of America}
\bvolume{108}(\bissue{2}),
\bfpage{729}--\blpage{741}
(\byear{2018})
\end{barticle}
\endbibitem

\bibitem[\protect\citeauthoryear{Rasmussen}{2013}]{rasmussen2013bayesian}
\begin{barticle}
\bauthor{\bsnm{Rasmussen}, \binits{J.G.}}:
\batitle{Bayesian inference for hawkes processes}.
\bjtitle{Methodology and Computing in Applied Probability}
\bvolume{15},
\bfpage{623}--\blpage{642}
(\byear{2013})
\end{barticle}
\endbibitem

\bibitem[\protect\citeauthoryear{Rasmussen}{2018}]{rasmussen2018lecture}
\begin{botherref}
\oauthor{\bsnm{Rasmussen}, \binits{J.G.}}:
Lecture notes: Temporal point processes and the conditional intensity function.
arXiv preprint arXiv:1806.00221
(2018)
\end{botherref}
\endbibitem

\bibitem[\protect\citeauthoryear{Rathbun}{1996}]{rathbun1996asymptotic}
\begin{barticle}
\bauthor{\bsnm{Rathbun}, \binits{S.L.}}:
\batitle{Asymptotic properties of the maximum likelihood estimator for spatio-temporal point processes}.
\bjtitle{Journal of Statistical Planning and Inference}
\bvolume{51}(\bissue{1}),
\bfpage{55}--\blpage{74}
(\byear{1996})
\end{barticle}
\endbibitem

\bibitem[\protect\citeauthoryear{Rhoades et~al.}{2018}]{rhoades2018highlights}
\begin{barticle}
\bauthor{\bsnm{Rhoades}, \binits{D.A.}},
\bauthor{\bsnm{Christophersen}, \binits{A.}},
\bauthor{\bsnm{Gerstenberger}, \binits{M.C.}},
\bauthor{\bsnm{Liukis}, \binits{M.}},
\bauthor{\bsnm{Silva}, \binits{F.}},
\bauthor{\bsnm{Marzocchi}, \binits{W.}},
\bauthor{\bsnm{Werner}, \binits{M.J.}},
\bauthor{\bsnm{Jordan}, \binits{T.H.}}:
\batitle{Highlights from the first ten years of the new zealand earthquake forecast testing center}.
\bjtitle{Seismological Research Letters}
\bvolume{89}(\bissue{4}),
\bfpage{1229}--\blpage{1237}
(\byear{2018})
\end{barticle}
\endbibitem

\bibitem[\protect\citeauthoryear{Reinhart}{2018}]{reinhart2018review}
\begin{barticle}
\bauthor{\bsnm{Reinhart}, \binits{A.}}:
\batitle{A review of self-exciting spatio-temporal point processes and their applications}.
\bjtitle{Statistical Science}
\bvolume{33}(\bissue{3}),
\bfpage{299}--\blpage{318}
(\byear{2018})
\end{barticle}
\endbibitem

\bibitem[\protect\citeauthoryear{Ripley}{1977}]{ripley1977modelling}
\begin{barticle}
\bauthor{\bsnm{Ripley}, \binits{B.D.}}:
\batitle{Modelling spatial patterns}.
\bjtitle{Journal of the Royal Statistical Society: Series B (Methodological)}
\bvolume{39}(\bissue{2}),
\bfpage{172}--\blpage{192}
(\byear{1977})
\end{barticle}
\endbibitem

\bibitem[\protect\citeauthoryear{Rhoades et~al.}{2016}]{rhoades2016retrospective}
\begin{barticle}
\bauthor{\bsnm{Rhoades}, \binits{D.}},
\bauthor{\bsnm{Liukis}, \binits{M.}},
\bauthor{\bsnm{Christophersen}, \binits{A.}},
\bauthor{\bsnm{Gerstenberger}, \binits{M.}}:
\batitle{Retrospective tests of hybrid operational earthquake forecasting models for canterbury}.
\bjtitle{Geophysical Journal International}
\bvolume{204}(\bissue{1}),
\bfpage{440}--\blpage{456}
(\byear{2016})
\end{barticle}
\endbibitem

\bibitem[\protect\citeauthoryear{Rezende and Mohamed}{2015}]{rezende2015variational}
\begin{bchapter}
\bauthor{\bsnm{Rezende}, \binits{D.}},
\bauthor{\bsnm{Mohamed}, \binits{S.}}:
\bctitle{Variational inference with normalizing flows}.
In: \bbtitle{International Conference on Machine Learning},
pp. \bfpage{1530}--\blpage{1538}
(\byear{2015}).
\bcomment{PMLR}
\end{bchapter}
\endbibitem

\bibitem[\protect\citeauthoryear{Ross}{2021}]{ross2021bayesian}
\begin{barticle}
\bauthor{\bsnm{Ross}, \binits{G.J.}}:
\batitle{Bayesian estimation of the etas model for earthquake occurrences}.
\bjtitle{Bulletin of the Seismological Society of America}
\bvolume{111}(\bissue{3}),
\bfpage{1473}--\blpage{1480}
(\byear{2021})
\end{barticle}
\endbibitem

\bibitem[\protect\citeauthoryear{Rue et~al.}{2017}]{rue2017bayesian}
\begin{barticle}
\bauthor{\bsnm{Rue}, \binits{H.}},
\bauthor{\bsnm{Riebler}, \binits{A.}},
\bauthor{\bsnm{S{\o}rbye}, \binits{S.H.}},
\bauthor{\bsnm{Illian}, \binits{J.B.}},
\bauthor{\bsnm{Simpson}, \binits{D.P.}},
\bauthor{\bsnm{Lindgren}, \binits{F.K.}}:
\batitle{Bayesian computing with inla: a review}.
\bjtitle{Annual Review of Statistics and Its Application}
\bvolume{4},
\bfpage{395}--\blpage{421}
(\byear{2017})
\end{barticle}
\endbibitem

\bibitem[\protect\citeauthoryear{Ross et~al.}{2019}]{ross2019searching}
\begin{barticle}
\bauthor{\bsnm{Ross}, \binits{Z.E.}},
\bauthor{\bsnm{Trugman}, \binits{D.T.}},
\bauthor{\bsnm{Hauksson}, \binits{E.}},
\bauthor{\bsnm{Shearer}, \binits{P.M.}}:
\batitle{Searching for hidden earthquakes in southern california}.
\bjtitle{Science}
\bvolume{364}(\bissue{6442}),
\bfpage{767}--\blpage{771}
(\byear{2019})
\end{barticle}
\endbibitem

\bibitem[\protect\citeauthoryear{Serafini et~al.}{2023}]{serafini2023approximation}
\begin{botherref}
\oauthor{\bsnm{Serafini}, \binits{F.}},
\oauthor{\bsnm{Lindgren}, \binits{F.}},
\oauthor{\bsnm{Naylor}, \binits{M.}}:
Approximation of bayesian hawkes process with inlabru.
Environmetrics,
2798
(2023)
\end{botherref}
\endbibitem

\bibitem[\protect\citeauthoryear{Stockman et~al.}{2023}]{stockman2023forecasting}
\begin{barticle}
\bauthor{\bsnm{Stockman}, \binits{S.}},
\bauthor{\bsnm{Lawson}, \binits{D.J.}},
\bauthor{\bsnm{Werner}, \binits{M.J.}}:
\batitle{Forecasting the 2016–2017 central apennines earthquake sequence with a neural point process}.
\bjtitle{Earth's Future}
\bvolume{11}(\bissue{9}),
\bfpage{2023}--\blpage{003777}
(\byear{2023})
\doiurl{10.1029/2023EF003777} .
\bcomment{e2023EF003777 2023EF003777}
\end{barticle}
\endbibitem

\bibitem[\protect\citeauthoryear{Seif et~al.}{2017}]{seif2017estimating}
\begin{barticle}
\bauthor{\bsnm{Seif}, \binits{S.}},
\bauthor{\bsnm{Mignan}, \binits{A.}},
\bauthor{\bsnm{Zechar}, \binits{J.D.}},
\bauthor{\bsnm{Werner}, \binits{M.J.}},
\bauthor{\bsnm{Wiemer}, \binits{S.}}:
\batitle{Estimating etas: The effects of truncation, missing data, and model assumptions}.
\bjtitle{Journal of Geophysical Research: Solid Earth}
\bvolume{122}(\bissue{1}),
\bfpage{449}--\blpage{469}
(\byear{2017})
\end{barticle}
\endbibitem

\bibitem[\protect\citeauthoryear{Sharrock et~al.}{2022}]{sharrock2022sequential}
\begin{botherref}
\oauthor{\bsnm{Sharrock}, \binits{L.}},
\oauthor{\bsnm{Simons}, \binits{J.}},
\oauthor{\bsnm{Liu}, \binits{S.}},
\oauthor{\bsnm{Beaumont}, \binits{M.}}:
Sequential neural score estimation: Likelihood-free inference with conditional score based diffusion models.
arXiv preprint arXiv:2210.04872
(2022)
\end{botherref}
\endbibitem

\bibitem[\protect\citeauthoryear{Sornette and Werner}{2005}]{sornette2005apparent}
\begin{botherref}
\oauthor{\bsnm{Sornette}, \binits{D.}},
\oauthor{\bsnm{Werner}, \binits{M.J.}}:
Apparent clustering and apparent background earthquakes biased by undetected seismicity.
Journal of Geophysical Research: Solid Earth
\textbf{110}(B9)
(2005)
\end{botherref}
\endbibitem

\bibitem[\protect\citeauthoryear{Taroni et~al.}{2018}]{taroni2018prospective}
\begin{barticle}
\bauthor{\bsnm{Taroni}, \binits{M.}},
\bauthor{\bsnm{Marzocchi}, \binits{W.}},
\bauthor{\bsnm{Schorlemmer}, \binits{D.}},
\bauthor{\bsnm{Werner}, \binits{M.J.}},
\bauthor{\bsnm{Wiemer}, \binits{S.}},
\bauthor{\bsnm{Zechar}, \binits{J.D.}},
\bauthor{\bsnm{Heiniger}, \binits{L.}},
\bauthor{\bsnm{Euchner}, \binits{F.}}:
\batitle{Prospective csep evaluation of 1-day, 3-month, and 5-yr earthquake forecasts for italy}.
\bjtitle{Seismological Research Letters}
\bvolume{89}(\bissue{4}),
\bfpage{1251}--\blpage{1261}
(\byear{2018})
\end{barticle}
\endbibitem

\bibitem[\protect\citeauthoryear{Utsu et~al.}{1995}]{utsu1995centenary}
\begin{barticle}
\bauthor{\bsnm{Utsu}, \binits{T.}},
\bauthor{\bsnm{Ogata}, \binits{Y.}}, \betal:
\batitle{The centenary of the omori formula for a decay law of aftershock activity}.
\bjtitle{Journal of Physics of the Earth}
\bvolume{43}(\bissue{1}),
\bfpage{1}--\blpage{33}
(\byear{1995})
\end{barticle}
\endbibitem

\bibitem[\protect\citeauthoryear{Utsu}{1970}]{utsu1970aftershocks}
\begin{barticle}
\bauthor{\bsnm{Utsu}, \binits{T.}}:
\batitle{Aftershocks and earthquake statistics (1): Some parameters which characterize an aftershock sequence and their interrelations}.
\bjtitle{Journal of the Faculty of Science, Hokkaido University. Series 7, Geophysics}
\bvolume{3}(\bissue{3}),
\bfpage{129}--\blpage{195}
(\byear{1970})
\end{barticle}
\endbibitem

\bibitem[\protect\citeauthoryear{Vargas and Gneiting}{2012}]{vargas2012bayesian}
\begin{botherref}
\oauthor{\bsnm{Vargas}, \binits{N.}},
\oauthor{\bsnm{Gneiting}, \binits{T.}}:
Bayesian point process modelling of earthquake occurrences.
Technical report,
Technical Report, Ruprecht-Karls University Heidelberg, Heidelberg, Germany~…
(2012)
\end{botherref}
\endbibitem

\bibitem[\protect\citeauthoryear{White et~al.}{2019}]{white2019detailed}
\begin{barticle}
\bauthor{\bsnm{White}, \binits{M.C.}},
\bauthor{\bsnm{Ben-Zion}, \binits{Y.}},
\bauthor{\bsnm{Vernon}, \binits{F.L.}}:
\batitle{A detailed earthquake catalog for the san jacinto fault-zone region in southern california}.
\bjtitle{Journal of Geophysical Research: Solid Earth}
\bvolume{124}(\bissue{7}),
\bfpage{6908}--\blpage{6930}
(\byear{2019})
\end{barticle}
\endbibitem

\bibitem[\protect\citeauthoryear{Wang et~al.}{2020}]{wang2020optimizing}
\begin{barticle}
\bauthor{\bsnm{Wang}, \binits{Y.}},
\bauthor{\bsnm{Gui}, \binits{Z.}},
\bauthor{\bsnm{Wu}, \binits{H.}},
\bauthor{\bsnm{Peng}, \binits{D.}},
\bauthor{\bsnm{Wu}, \binits{J.}},
\bauthor{\bsnm{Cui}, \binits{Z.}}:
\batitle{Optimizing and accelerating space--time ripley’s k function based on apache spark for distributed spatiotemporal point pattern analysis}.
\bjtitle{Future Generation Computer Systems}
\bvolume{105},
\bfpage{96}--\blpage{118}
(\byear{2020})
\end{barticle}
\endbibitem

\bibitem[\protect\citeauthoryear{Woessner et~al.}{2011}]{woessner2011retrospective}
\begin{botherref}
\oauthor{\bsnm{Woessner}, \binits{J.}},
\oauthor{\bsnm{Hainzl}, \binits{S.}},
\oauthor{\bsnm{Marzocchi}, \binits{W.}},
\oauthor{\bsnm{Werner}, \binits{M.}},
\oauthor{\bsnm{Lombardi}, \binits{A.}},
\oauthor{\bsnm{Catalli}, \binits{F.}},
\oauthor{\bsnm{Enescu}, \binits{B.}},
\oauthor{\bsnm{Cocco}, \binits{M.}},
\oauthor{\bsnm{Gerstenberger}, \binits{M.}},
\oauthor{\bsnm{Wiemer}, \binits{S.}}:
A retrospective comparative forecast test on the 1992 landers sequence.
Journal of Geophysical Research: Solid Earth
\textbf{116}(B5)
(2011)
\end{botherref}
\endbibitem

\bibitem[\protect\citeauthoryear{Wang et~al.}{2010}]{wang2010standard}
\begin{barticle}
\bauthor{\bsnm{Wang}, \binits{Q.}},
\bauthor{\bsnm{Schoenberg}, \binits{F.P.}},
\bauthor{\bsnm{Jackson}, \binits{D.D.}}:
\batitle{Standard errors of parameter estimates in the etas model}.
\bjtitle{Bulletin of the Seismological Society of America}
\bvolume{100}(\bissue{5A}),
\bfpage{1989}--\blpage{2001}
(\byear{2010})
\end{barticle}
\endbibitem

\bibitem[\protect\citeauthoryear{Xing et~al.}{2019}]{xing2019calibrated}
\begin{bchapter}
\bauthor{\bsnm{Xing}, \binits{H.}},
\bauthor{\bsnm{Nicholls}, \binits{G.}},
\bauthor{\bsnm{Lee}, \binits{J.}}:
\bctitle{Calibrated approximate bayesian inference}.
In: \bbtitle{International Conference on Machine Learning},
pp. \bfpage{6912}--\blpage{6920}
(\byear{2019}).
\bcomment{PMLR}
\end{bchapter}
\endbibitem

\bibitem[\protect\citeauthoryear{Zhu and Beroza}{2019}]{zhu2019phasenet}
\begin{barticle}
\bauthor{\bsnm{Zhu}, \binits{W.}},
\bauthor{\bsnm{Beroza}, \binits{G.C.}}:
\batitle{Phasenet: a deep-neural-network-based seismic arrival-time picking method}.
\bjtitle{Geophysical Journal International}
\bvolume{216}(\bissue{1}),
\bfpage{261}--\blpage{273}
(\byear{2019})
\end{barticle}
\endbibitem

\bibitem[\protect\citeauthoryear{Zhuang et~al.}{2012}]{zhuang2012basic}
\begin{botherref}
\oauthor{\bsnm{Zhuang}, \binits{J.}},
\oauthor{\bsnm{Harte}, \binits{D.S.}},
\oauthor{\bsnm{Werner}, \binits{M.J.}},
\oauthor{\bsnm{Hainzl}, \binits{S.}},
\oauthor{\bsnm{Zhou}, \binits{S.}}:
Basic models of seismicity: Temporal models
(2012)
\end{botherref}
\endbibitem

\bibitem[\protect\citeauthoryear{Zhuang et~al.}{2004}]{zhuang2004analyzing}
\begin{botherref}
\oauthor{\bsnm{Zhuang}, \binits{J.}},
\oauthor{\bsnm{Ogata}, \binits{Y.}},
\oauthor{\bsnm{Vere-Jones}, \binits{D.}}:
Analyzing earthquake clustering features by using stochastic reconstruction.
Journal of Geophysical Research: Solid Earth
\textbf{109}(B5)
(2004)
\end{botherref}
\endbibitem

\bibitem[\protect\citeauthoryear{Zhuang et~al.}{2017}]{zhuang2017data}
\begin{barticle}
\bauthor{\bsnm{Zhuang}, \binits{J.}},
\bauthor{\bsnm{Ogata}, \binits{Y.}},
\bauthor{\bsnm{Wang}, \binits{T.}}:
\batitle{Data completeness of the kumamoto earthquake sequence in the jma catalog and its influence on the estimation of the etas parameters}.
\bjtitle{Earth, Planets and Space}
\bvolume{69}(\bissue{1}),
\bfpage{1}--\blpage{12}
(\byear{2017})
\end{barticle}
\endbibitem

\end{thebibliography}

\end{document}